\documentclass[a4paper,12pt]{article}
\usepackage{graphicx}
\usepackage{amsmath}
\usepackage{amssymb}
\usepackage{latexsym}
\usepackage{psfrag}
\newsavebox{\LSIM}
\sbox{\LSIM}{\raisebox{-1ex}{$\ \stackrel{\textstyle<}{\sim}\ $}}

\newsavebox{\GSIM}
\sbox{\GSIM}{\raisebox{-1ex}{$\ \stackrel{\textstyle>}{\sim}\ $}}

\begin{document}
\begin{titlepage}
\begin{flushright}
\end{flushright}
$\mbox{ }$
\vspace{.1cm}
\begin{center}
\vspace{.5cm}
{\bf\Large An efficient approach to electroweak bubble velocities }\\[.3cm]
\vspace{1cm}
Stephan J.~Huber$^{a,}$\footnote{s.huber@sussex.ac.uk}
and 
Miguel Sopena$^{a,}$\footnote{m.sopena@sussex.ac.uk} \\ 
\vspace{1cm} {\em  
$^a$ Department of Physics and Astronomy, Sussex University, Brighton, East Sussex BN1 9QE, UK}\\[.2cm] 
\end{center}
\vspace{1.cm}
\begin{abstract}
Extensions of the Standard Model are being considered as viable settings for a first-order electroweak phase transition which satisfy Sakharov's three conditions for the generation of the baryon asymmetry of the Universe. These extensions  provide a sufficiently strong phase transition and remove the main obstacles which appear in the context of the Standard Model: A far-too-high lower bound on the Higgs mass, immediate wipeout of the newly-created baryon asymmetry, and insufficient CP violation.
We describe the Universe hydrodynamically as a fluid coupled to the Higgs field via a phenomenological friction term, and study the time evolution of bubbles nucleated during the phase transition. We express the friction term in the hydrodynamic equations in terms of the particle content of the model, calibrate the friction on the basis of existing calculations for the Standard Model, and produce predictions for the velocity of the expanding bubble wall in the stationary regime. This way we develop a very efficient approach to compute bubble velocities.
As an example, we apply our formalism to the  first-order phase transition of a dimension-6 extension of the Standard Model which, within the present bounds on the Higgs mass, can reproduce the observed baryon asymmetry of the Universe.  Depending on the strength of the phase transition, the wall velocity varies from about 0.3 to approaching the speed of light. Our method can easily be adapted to compute wall velocities in other interesting extensions of the Standard Model.

\end{abstract}
\end{titlepage}

\section{Introduction}
A first-order electroweak phase transition (EWPT) in which two differentiated phases briefly coexisted could be the setting for the genesis of the observed baryon asymmetry of the Universe \cite{Sakharov:1967dj,'tHooft:1976up,Kuzmin:1985mm,Anderson:1991zb} (for a review, see e.g.~\cite{Cline:2006ts}), and may have left signatures, like a gravitational wave background or primordial magnetic fields, which we might be able to observe in relatively short order  \cite{Grasso:2000wj,Vilenkin:1986hg,Megevand:2004ry,Grojean:2006bp,Huber:2007vva,Huber:2008hg}. Such a transition may satisfy Andrei Sakharov's three conditions for electroweak baryogenesis \cite{Sakharov:1967dj}: CP violation, baryon number violation, and deviation from thermal equilibrium, the latter imposed by the expanding bubbles of the broken symmetry phase on the whole of the medium at one point or another until the time when the whole Universe has crossed over to the new phase and the transition ends. A first-order EWPT requires the presence of two stable or metastable phases which correspond to extrema in the high-temperature expansion of the Higgs thermal effective potential (Figure \ref{fig:effectivepotential}). It is well known that the observed baryon asymmetry could not have been produced in such a scenario within the Standard Model (SM) unless the Higgs mass were far lower than allowed by present experimental bounds\footnote{The Atlas and CMS experiments at the Large Hadron Collider (LHC) had bracketed the Higgs mass to between $117$ and $127$ GeV when the discovery of a 'previously unknown boson', consistent with the Higgs and with a mass of $125-127$ GeV, was announced in July 2012 
\cite{NatureDec2011,:2012gu,:2012gk}}. This is because sphaleron transitions must be sufficiently suppressed inside the bubbles of the new phase to avoid washout of the newly-created baryon asymmetry \cite{Khlebnikov:1988sr,Bodeker:1999gx,Shaposhnikov:1986jp}. Also, the electroweak phase transition is not even first-order in the Standard Model except (again) for Higgs masses far below present experimental bounds \cite{Kajantie:1996mn}. Simple extensions of the standard model, however, are still capable of producing a sufficiently high baryon asymmetry through a first-order EWPT for presently acceptable values of the Higgs mass $m_H$, be it via
extra Higgs fields \cite{McLerran:1988bb,Fromme:2006cm,Cline:2011mm}, light top squarks \cite{Carena:1996wj,Bodeker:1996pc,Laine:1998qk,Carena:2008vj},  singlets \cite{Anderson:1991zb,Pietroni:1992in,Huber:2000mg,Menon:2004wv,Huber:2006ma,Espinosa:2011ax,Espinosa:2011eu} or non-standard Higgs potentials \cite{Grojean:2004xa,Bodeker:2004ws}. Below we will consider the latter possibility.

Calculations of the dynamical characteristics of a first-order phase transition (like the expansion velocity of the walls of bubbles of the new phase after nucleation) mostly rely on solving the relevant hydrodynamical equations, modelling the radiation-dominated early universe as a perfect relativistic fluid \cite{Megevand:2009gh,Ignatius:1993qn,Moore:1995si,Huet:1992ex,John:2000zq,Bodeker:2009qy,Megevand:2009ut}. Usually, the WKB (semiclassical) approximation is adopted. This assumes that the de Broglie wavelength of the particles in the plasma is considerably smaller than the width of the bubble wall, which should be true for all particles in most situations except possibly for very infrared bosons, which are assumed to have little influence on the wall dynamics \cite{Moore:1995si}. Such treatments derive a fluid equation which describes the dynamics of the variables characterising the cosmic fluid (fluid velocity, temperature, and Higgs vacuum expectation value). The fluid equation includes a \textit{friction term} which tends to slow down the propagation of the bubble wall. In some treatments a phenomenological parameter which characterises the medium is introduced into such a term \cite{Ignatius:1993qn,Megevand:2009ut}. In others an explicit calculation of the friction from the modelling of the plasma is carried out \cite{Moore:1995si,John:2000zq}. Friction is a function of the deviation from thermal equilibrium of the particle components of the plasma which couple strongly to the background Higgs field. If local thermal equilibrium (LTE) were maintained throughout the plasma there would be no friction and the expanding bubble walls would accelerate to the speed of light \cite{Turok:1992jp}. The presence of friction may lead to two outcomes: Either the advancing bubble wall reaches a steady state under the influence of friction and propagates with constant velocity, or friction is too weak and the wall accelerates without bound towards ultrarelativistic velocities. The feasibility of such 'runaway walls' has recently been proven theoretically \cite{Grojean:2004xa,Bodeker:2009qy}.

In the following we lay out a general approach to bubble velocities, based on an efficient parametrization of friction exerted by the plasma on the bubble wall. In ref.~\cite{Huber:2011aa}, we have used this framework for the first time to compute the wall velocity in the MSSM for realistic Higgs masses, requiring use of the 2-loop thermal Higgs potential. Here we take the example of the dimension-6 extension of the Standard Model with a cut-off $M$, discussed in detail in refs.~\cite{Bodeker:2004ws}, to demonstrate the power of our approach to bubble velocities. This extension if the SM provides also new sources of CP-violation necessary to obtain realistic values of the baryon asymmetry within the model parameters \cite{Huber:2006ri} and results in a phase transition strong enough to avoid washout. We apply the semiclassical, perfect-fluid approximation to the electroweak plasma with our extended thermal potential. 

We adapt standard hydrodynamical calculations to our model which employ a phenomenological friction parameter. In order to calibrate this parameter realistically for our model we base ourselves on existing results for the wall velocity within the Standard Model and the description of the evolution of particle populations through Boltzmann evolution equations in the form known as the relaxation time approximation \cite{Moore:1995si}. We compare the calibration from Standard Model results with that found through the theoretical criterion for 'runaway walls' mentioned above. Thus we predict values of the friction parameter with which the steady-state expansion velocity of the bubble walls may be calculated for a range of the model parameters $M$, $m_H$. 

We find the wall velocity to be often subsonic (lower than the speed of sound in the medium), similar to existing results for the minimal SM. For relatively low values of the Higgs mass and the dimension-6 cutoff scale $M$, however, the phase transition becomes very strong and the steady-state wall velocity may shift into the supersonic regime. Eventually we predict runaway, rather than steady-state, behaviour for the highest values of the strength of the phase transition within our parameter space. We comment briefly on the possibility of 'hybrid' steady-state solutions in the frontier between the subsonic and supersonic regimes, as well as on the existing criteria for stability of the growing bubbles as they apply to our model. 

Lastly, we provide a 'user-friendly' description of our friction parameter calibration, based on the relaxation time approximation, for use in any model given a reference calibration point for the friction. The results presented here can directly be used to treat cases with SM-like friction, e.g.~SUSY scenarios without light stops, as for instance can be realised in singlet extensions of the MSSM.

\section{Single field hydrodynamics and the electroweak phase transition}

\subsection{The effective potential and critical free energy}

\begin{figure}
\begin{center}
\includegraphics{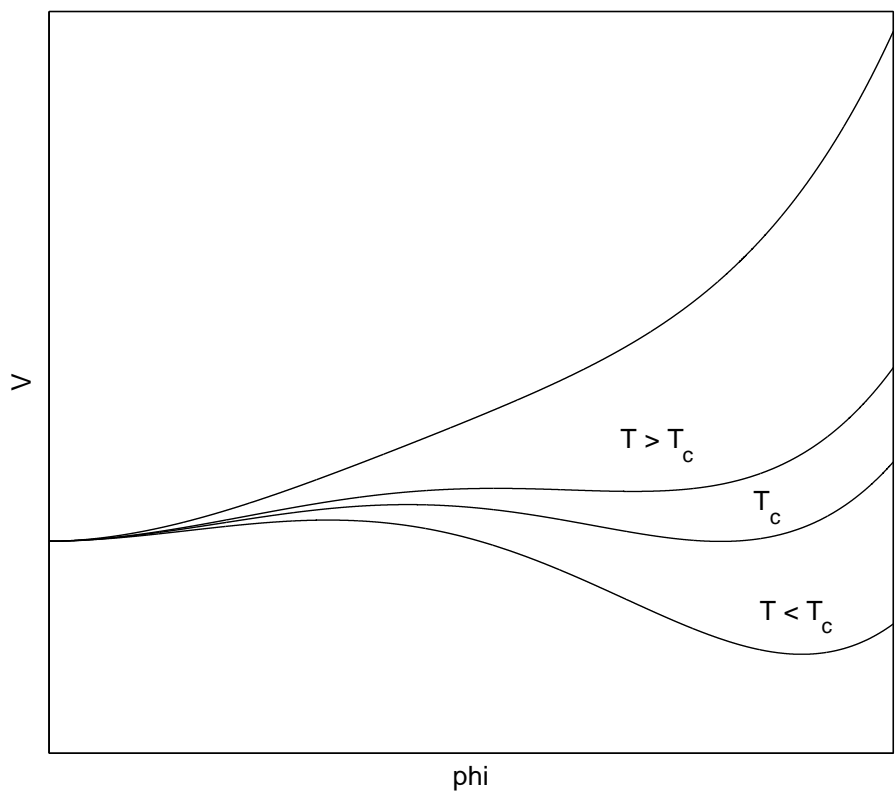}
\caption{Finite-temperature effective potential for a first-order phase transition.}
\label{fig:effectivepotential}
\end{center}
\end{figure}

The fundamental tool employed to study the electroweak phase transition is the finite-temperature effective potential for the background Higgs VEV $\phi$, $V(\phi, T)$, representing the free energy per unit volume of the field configuration (see e.g.~\cite{Anderson:1991zb}). At high temperature, the thermal potential has only one local minimum (Figure \ref{fig:effectivepotential}) at $\phi = 0$, representing the unbroken symmetry phase of the theory and the state of the universe before the phase transition. As temperature decreases a second local minimum of the potential appears for a $\phi_0 > 0$, corresponding to the broken symmetry phase of the theory. As at first $V(\phi_0,T) > V(0,T)$ the broken symmetry phase is energetically disfavoured and the phase transition may not yet proceed. As temperature decreases further, $V(\phi_0,T)$ also decreases until it becomes degenerate with the $0$-VEV minimum at the \textit{critical temperature} $T_c$. From that point on configurations ('bubbles') of the new phase may, in principle, form and the transition may take place. 

The free energy of a bubble of the broken symmetry phase is given by

\begin{equation}\label{eqn:Free energy}
	F=\int_{V}{d}^{3}x[\frac{1}{2}(\nabla\phi)^{2}+V(\phi)].
\end{equation}
The first term represents the positive contribution from the bubble wall, defined as the region where the Higgs VEV varies between its zero value in the symmetric phase outside the bubble and its nonzero value inside. The second term represents the negative contribution from the value of the Higgs potential in the interior volume of the bubble. Assuming spherically symmetric bubbles, for each temperature there exists a \textit{critical radius} such that nucleated bubbles larger than the critical size will expand spontaneously (thus minimising their free energy) whereas smaller than critical bubbles will collapse under surface tension. The critical radius is infinite at $T_c$ and decreases monotonically with temperature. The critical bubble is a {\em static} solution of the Euler-Lagrange equation

\begin{eqnarray}
	\partial_{\mu}\frac{\delta{\textit{L}}}{\delta\partial_{\mu}\phi}-\frac{\delta{\textit{L}}}{\delta\phi}=0.
\end{eqnarray}
In spherical coordinates with radial coordinate $r$ (and since for a static solution $\partial_{t}\equiv0$) the Euler-Lagrange equation becomes

\begin{eqnarray}
	\frac{d^{2}\phi}{dr^{2}}+\frac{2}{r}\frac{d\phi}{dr}=\frac{\partial{V}}{\partial\phi}\label{eqn:circularbubble}
\end{eqnarray}
which has to be solved with boundary conditions

\begin{eqnarray}
	\phi(+\infty)=0\nonumber\\
	\left[\frac{d\phi}{dr}\right]_{r=0}=0\nonumber
\end{eqnarray}

\begin{figure}
	\centering
		\includegraphics{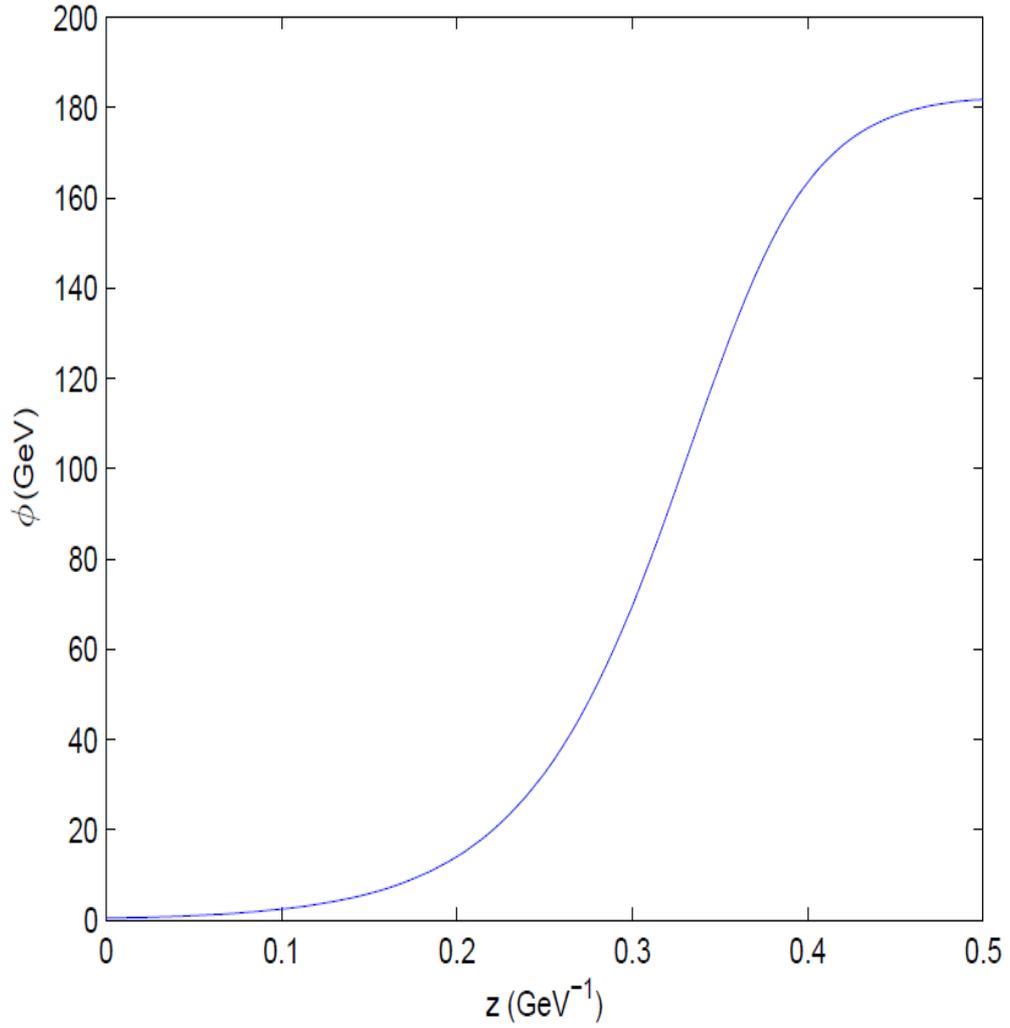}
	\caption{Critical bubble solution for the $\phi^6$ model with $M=800$ GeV, $m_h=115$ GeV,  $T=105.62$ GeV (for parameter definitions see section 5).}
	\label{fig:critbub}
\end{figure}

The critical bubble solution can be integrated numerically (eq (\ref{eqn:Free energy})) over the bubble volume to find the critical free energy. The radius of the critical bubble at the critical temperature becomes infinite and the equation for the Higgs VEV may be written as

\begin{eqnarray}
	\frac{d^{2}\phi}{dz^{2}}=\frac{\partial{V}}{\partial\phi}\label{eqn:planarwall}
\end{eqnarray}
with boundary conditions

\begin{eqnarray}
	\phi(-\infty)=0\nonumber\\
	\phi(+\infty)={v}_c\nonumber
\end{eqnarray}
where $v_c$ is the nonzero Higgs VEV at the degenerate minimum at $T=T_c$ (Fig \ref{fig:critbub}).

\subsection{Nucleation and finalisation temperatures}

The probability of bubble nucleation within the horizon volume in the time interval $dt$ may be calculated from the critical free energy as (see again e.g.~\cite{Anderson:1991zb})

\begin{eqnarray}
	dP=\left(\Gamma/\textit{V}\right)\cdot{V}_H\cdot{dt}=\frac{T^4}{H^4}e^{-F_{c}/T}\frac{dT}{T}\nonumber
\end{eqnarray}
where $\Gamma/\textit{V}=\Lambda^{4}(T)e^{-F_{c}/T}\approx T^4 e^{-F_{c}/T}$ is the nucleation rate per unit volume with $F_c$ the critical free energy at the temperature $T$. The horizon volume is $V_H=d_H^3=(2t)^3=H^{-3}$ with $H^{-1}=\frac{2\xi{M}_{Pl}}{T^{2}}$ and $M_{Pl}$ the Planck mass. $\xi\cong(1/34)$ at the time of the electroweak phase transition. The phase transition starts at the \textit{nucleation temperature} for which

\begin{eqnarray}
	P(T=T_n)=\int^{T_c}_{T_n}dP\equiv1\label{eqn:nuctemp}
\end{eqnarray}

A true-vacuum (broken phase) bubble nucleated at a temperature $T'$ and expanding with constant velocity $\beta$ will have grown to a radius\footnote{We ignore the effect of the expansion of the Universe, since the transition is extremely fast, so that $dR=\beta{dt}$.} at a later temperature $T$

\begin{eqnarray}
	R_B(T,T')=2\beta\xi\frac{M_{Pl}}{T}\left(\frac{1}{T}-\frac{1}{T'}\right)
\end{eqnarray}

Thus the fraction $f$ of the causal volume which has crossed over to the broken phase at a temperature $T<T_n$ is given by (neglecting mergers between bubbles)

\begin{eqnarray}
	f(T)&=&\frac{1}{V_H}\frac{4\pi}{3}\int^{T_c}_{T}R_B(T,T')^3dP(T')=\nonumber\\	&=&\frac{4\pi{H(T)}^{3}}{3}\int^{T_c}_{T}\left(2\beta\xi\frac{M_{Pl}}{T}\right)^3\left(\frac{1}{T}-\frac{1}{T'}\right)^3\frac{T'^4}{H(T')^4}e^{-F_{c}/T'}\frac{dT'}{T'}=\nonumber\\	
	&=&\frac{4\pi\beta^{3}}{3}\left(2\xi{M}_{Pl}\right)^4\int^{T_c}_{T}\left(1-\frac{T}{T'}\right)^3\frac{1}{T'^5}e^{-F_{c}/T'}dT'.\label{eqn:fraction}
\end{eqnarray}
$f(T_f)=1$ defines the phase transition's \textit{finalisation temperature} at which the transition can be assumed to end.

\subsection{Equation of motion for the Higgs VEV}

The equation of motion for the background Higgs field (see e.g.~\cite{Moore:1995si}) is a function of the phase space population density functions $f(p,x)$ for all the particles present in the plasma,
\begin{eqnarray} \label{eq:eomMP1}
\Box \phi+\frac{\partial V(\phi)}{\partial \phi}+\sum \frac{dm^2}{d\phi}\int\frac{d^3p}{(2\pi)^32E}f(p,x)=0
\end{eqnarray}
where $V(\phi)$ is the renormalised vacuum potential and the sum is over massive degrees of freedom so that, for the appropriate couplings, e.g.~$m^2=\frac{y^2\phi^2}{2}$ for fermions, $m^2=\frac{g^2\phi^2}{4}$ for bosons. The population densities can be expressed as a thermal equilibrium part plus a deviation, $f \equiv f_0 + \delta f$, with the equilibrium distribution for fermions/bosons, in the rest frame of the plasma, given by $f_{\rm{0}}(T)=\frac{1}{e^{E/T}\pm1}$ (with $E=\sqrt{m^2+\left|\vec{p}\right|^2}$). The derivative of the vacuum potential combines with the integral of the equilibrium part of the distributions to give the finite-temperature effective potential $V(\phi,T)$. The equation of motion then becomes
\begin{equation} \label{eq:eomMP2}
\Box \phi+\frac{\partial V(\phi,T)}{\partial \phi}+\sum\frac{dm^2}{d\phi}\int\frac{d^3p}{(2\pi)^32E}\delta f(p,x)=0.
\end{equation}
The $\delta f$-dependent term is the friction term. Thus, as stated earlier, friction effects are the result of the deviation of the particle populations in the plasma from equilibrium. If these did not exist there would be no friction (and the bubble wall would propagate at the speed of light). 

\subsection{The pressure on the wall}

A direct application of eq (\ref{eq:eomMP2}) is the calculation of the pressure felt by the wall \cite{Moore:1995si}. Assuming a planar wall advancing with a steady velocity in the $z$ direction we may rewrite (\ref{eq:eomMP2}) in the rest frame of the advancing wall as
\begin{equation} \label{eq:eomMP4}
\frac{d^2\phi}{dz^2} + \frac{\partial V(\phi,T)}{\partial \phi} + \sum\frac{dm^2}{d\phi}\int\frac{d^3p}{(2\pi)^32E}\delta f(p,x) = 0.
\end{equation}
Multiplying the whole expression by $\frac{d\phi}{dz} \equiv \phi'$ and integrating over $z$ across the wall, we find the pressure on the wall. Note that at both ends of the wall $\frac{d\phi}{dz} \equiv 0$. Therefore $\int \phi'' \phi' dz = \left[\frac{\phi'^2}{2}\right]^{\phi_0}_0 = 0$. We use the fact that the finite-temperature effective potential of the Higgs field equals the free energy density per unit volume $\textit{F}$, which is equal to minus the pressure. Thus integrating along the $z$-direction yields
\begin{equation} \label{eq:pressure}
\Delta V(\phi,T) + \int \phi' dz \sum \frac{dm^2}{d\phi}\int\frac{d^3p}{(2\pi)^32E}\delta f(p,x) = 0.
\end{equation}
In a steady state, the pressure felt by the advancing wall equalises with the integral of the friction term as written in eq.~(\ref{eq:pressure}).

\subsection{Boltzmann evolution equations}

In order to solve the equation of motion for the Higgs field, $\delta f$ must be expressed in an appropriate form. Within the WKB approximation the evolution of the particle populations is usually accounted for through the relevant Boltzmann evolution equations,
\begin{equation}
\frac{df}{dt}=\partial_tf+\dot{\vec{x}}\cdot\partial_{\vec{x}}f+\dot{\vec{p}}\cdot\partial_{\vec{p}}f=-C[f]
\end{equation}
where $C[f]$ is the so-called \textit{collision integral}, written in its full form as \cite{Moore:1995si}

\begin{eqnarray}
	C\;[f(x,p)] = \sum_i \frac{1}{2E_p} \int \frac{d^3k\;d^3p'\;d^3k'}{(2\pi)^9\;2E_k\;2E_{p'}\;2E_{k'}} \left|\textit{M}(s,t)\right|^2\nonumber\\
  \times (2 \pi)^2 \;  \delta^4(p+k-p'-k') \textit{P}\;[\textit{f}_\textit{i}]
\end{eqnarray}
and $P[\textit{f}_\textit{i}] = f_1 f_2 (1 \pm f_3)(1 \pm f_4) - f_3 f_4 (1 \pm f_1)(1 \pm f_2)$. The sum is over all relevant four leg scattering diagrams. \textit{p} and \textit{k} are the incoming, \textit{p'} and \textit{k'} the outgoing momenta. \textit{M} is the scattering amplitude for each process. The $f_i$ are population factors. The first (positive) contribution to \textit{P} represents a particle with momentum \textit{p} being removed from the state by a collision, weighted by the populations of colliding particles. The second (negative) contribution represents a particle being scattered into the state. The $(1 \pm f)$ factors stem from particle statistics ($-$ for fermions and $+$ for bosons).

The equilibrium distribution for fermions/bosons can be generalised to a frame in which the fluid is not at rest but has a bulk velocity $v$ in, say, the $z$ direction. The equilibrium distribution becomes
\begin{equation}
f_0=\frac{1}{e^{\beta \gamma (E-vp_z)}\pm1}
\end{equation}
with $\beta \equiv \frac{1}{T}$ and $\gamma=\frac{1}{\sqrt{1-v^2}}$. Taking $v = 0$ recovers the version for a fluid at rest. In order to work with the Boltzmann equations we need the relevant position and momentum derivatives of the equilibrium distributions,
\begin{equation}
\frac{df_0}{dp_z}=-\frac{\beta \gamma (\frac{p_z}{E}-v) e^{\beta \gamma (E-vp_z)}}{(e^{\beta \gamma (E-vp_z)}\pm1)^2}
\end{equation}
and
\begin{equation}
\frac{df_0}{dz}=-\frac{\beta \gamma \frac{(m^2)'}{2E} e^{\beta \gamma (E-vp_z)}}{(e^{\beta \gamma (E-vp_z)}\pm1)^2}.
\end{equation}

The hydrodynamical calculations across the bubble wall profile are usually carried out in the rest frame of the (steady-state) bubble wall, in which $\partial_t \equiv 0$ for all quantities. A planar wall will 'see' the plasma as moving with a given 'fluid' velocity. Note that, if the plasma itself has a position-dependent, nonzero bulk motion, induced by the passing of the wall, the plasma velocity 'seen' by the wall at any given position may not coincide with the steady-state velocity of wall propagation. Taking $\partial_{\vec{p}} \delta f \equiv 0$, and noting that $v_z\cdot\partial_z\delta f=\frac{p_z}{E}\cdot\partial_z\delta f$ and $\dot{\vec{p_z}}=-\frac{\partial E}{\partial z}\vec{u_z}=-\frac{1}{2E}\frac{d(m^2)}{dz}\vec{u_z}$, the Boltzmann equation in the rest frame of the advancing wall becomes
\begin{eqnarray}
\partial_t f+\dot{\vec{x}}\cdot\partial_{\vec{x}} f+\dot{\vec{p}}\cdot\partial_{\vec{p}} f=
	\nonumber\\
\frac{p_z}{E} \partial_z (f_0+\delta f)-\frac{(m^2)'}{2E}\partial_{p_z} f_0=
	\nonumber\\
\left[\frac{(m^2)'}{2E}(\frac{p_z}{E} - v) - \frac{p_z}{E} \frac{(m^2)'}{2E}\right]  \beta \gamma \frac{e^{\beta \gamma (E-v p_z)}}{(e^{\beta \gamma (E-vp_z)}\pm1)^2}+ \frac{p_z}{E} \partial_z \delta f=
	\nonumber\\
-\frac{(m^2)'}{2E}v \beta \gamma \frac{e^{\beta \gamma (E-v p_z)}}{(e^{\beta \gamma (E-vp_z)}\pm1)^2}+ \frac{p_z}{E} \delta f'=-C[f].\label{eq:Boltzboosted}
\end{eqnarray}

\subsection{The relaxation time approximation}

A number of simplifications are possible for the collision integral $C[f]$. The \textit{free particle approximation} \cite{Dine:1992wr,Liu:1992tn} assumes $C[f] \equiv 0$. This may represent the case in which particle free paths are much larger than the thickness of the wall. In this approximation the Boltzmann equations can be solved exactly (see e.g.~\cite{Moore:1995si}). 

Our starting point will be the \textit{relaxation time approximation} in which we assume $C[f] \equiv \frac{\delta f}{\tau}$, where the relaxation timescale $\tau$ is usually considered independent of momentum \cite{Dine:1992wr}. Following this approximation the Boltzmann equation becomes
\begin{eqnarray}
-\frac{(m^2)'}{2E}v \beta \gamma \frac{e^{\beta \gamma (E-v p_z)}}{(e^{\beta \gamma (E-vp_z)}\pm1)^2}+ \frac{p_z}{E} \delta f' = -\frac{\delta f}{\tau}.\label{eq:Boltzboosted2}
\end{eqnarray}
The usual procedure is to neglect either the $\delta f$ or the $\delta f'$ term. The authors of e.g.~\cite{Moore:1995si} assume $L \gg \tau$ for the wall thickness $L$, so that $\delta f' \approx \frac{\delta f}{L} \ll \frac{\delta f}{\tau}$. This is a natural assumption if the bubble wall is expected to be relatively \textit{slow} since slower walls are thicker and not as sharp (quantities vary more slowly across the wall) and relativistic time dilation is not a factor in increasing the characteristic timescale for particle interaction (the relaxation time). In this approximation we drop the derivative term in (\ref{eq:Boltzboosted}) and obtain (for one degree of freedom),
\begin{eqnarray}
\delta f_{\rm{slow\:wall}} = \tau \frac{(m^2)'}{2E}v \beta \gamma \frac{e^{\beta \gamma (E-v p_z)}}{(e^{\beta \gamma (E-vp_z)}\pm1)^2}\label{eq:deltafslowwall}
\end{eqnarray}
Inserting this form for $\delta f$ into the friction term in the equation of motion for the background Higgs field (\ref{eq:eomMP4}) we obtain
\begin{eqnarray}\label{eq:fricslowwall}
	\frac{dm^2}{d\phi}\int\frac{d^3p}{(2\pi)^32E}\delta f_{\rm{sw}}(p,x)=
	\nonumber\\
	\frac{dm^2}{d\phi}\int\frac{d^3p}{(2\pi)^32E}\tau \beta \gamma v \frac{(m^2)'}{2E}\frac{e^{\beta \gamma (E-v p_z)}}{(e^{\beta \gamma (E-vp_z)}\pm1)^2}
\end{eqnarray}
which gives an equation of motion for a slow wall, in the rest frame of the advancing wall, of the form
\begin{equation} \label{eq:eomsw}
\frac{d^2\phi}{dz^2} + \frac{\partial V(\phi,T)}{\partial \phi} + \sum\frac{dm^2}{d\phi}\int\frac{d^3p}{(2\pi)^32E}\tau \beta \gamma v \frac{(m^2)'}{2E}\frac{e^{\beta \gamma (E-v p_z)}}{(e^{\beta \gamma (E-vp_z)}\pm1)^2}= 0.
\end{equation}
The pressure difference felt by the advancing wall (for one degree of freedom) in this approximation is
\begin{eqnarray}\label{eq:deltaPslowwall}
	\Delta P_{\rm{sw}} = \int \phi' dz \frac{dm^2}{d\phi}\int\frac{d^3p}{(2\pi)^3 2E}\tau \beta \gamma v \frac{(m^2)'}{2E}\frac{e^{\beta \gamma (E-v p_z)}}{(e^{\beta \gamma (E-vp_z)}\pm1)^2}.
\end{eqnarray}

The opposite limit is that of a fast, relativistic wall. This will naturally be a thin, sharp wall and relativistic time dilation will lenghten the characteristic relaxation time for particle populations. Thus we assume $L \ll \tau$ and $\delta f' \approx \frac{\delta f}{L} \gg \frac{\delta f}{\tau}$. Therefore in this limit we may neglect the $\delta f$ term in (\ref{eq:Boltzboosted2}) and obtain
\begin{eqnarray}
	\frac{p_z}{E} \delta f' = \frac{(m^2)'}{2E} v \beta \gamma \frac{e^{\beta \gamma (E-v p_z)}}{(e^{\beta \gamma (E-vp_z)}\pm1)^2}\label{eq:Boltzboosted4}
\end{eqnarray}
which (assuming for simplicity $E \approx p_z$ in the exponentials) can be integrated to
\begin{eqnarray}
	\delta f_{\rm{fast\:wall}} = \frac{m^2}{2 P_z} v \beta \gamma \frac{e^{\beta \gamma (E-v p_z)}}{(e^{\beta \gamma (E-vp_z)}\pm1)^2}.\label{eq:Boltzboosted5}
\end{eqnarray}
The friction term becomes in this limit
\begin{eqnarray}
	\frac{dm^2}{d\phi}\int\frac{d^3p}{(2\pi)^32E}\delta f_{\rm{fw}}(p,x)=
	\nonumber\\
	\frac{dm^2}{d\phi}\int\frac{d^3p}{(2\pi)^32E} \frac{m^2}{2 P_z} v \beta \gamma \frac{e^{\beta \gamma (E-v p_z)}}{(e^{\beta \gamma (E-vp_z)}\pm1)^2}
	\nonumber\\
\end{eqnarray}
and the equation of motion for the fast wall case
\begin{equation} \label{eq:eomfw}
\frac{d^2\phi}{dz^2} + \frac{\partial V(\phi,T)}{\partial \phi} + \sum\frac{dm^2}{d\phi}\int\frac{d^3p}{(2\pi)^32E} \frac{m^2}{2 p_z} v \beta \gamma \frac{e^{\beta \gamma (E-v p_z)}}{(e^{\beta \gamma (E-vp_z)}\pm1)^2}= 0.
\end{equation}
The pressure on the wall (for one degree of freedom) in the fast wall limit is therefore
\begin{eqnarray}\label{eq:deltaPfastwall}
	\Delta P_{\rm{fw}} = \int \phi' dz 	\frac{dm^2}{d\phi}\int\frac{d^3p}{(2\pi)^32E} \frac{m^2}{2 p_z} v \beta \gamma \frac{e^{\beta \gamma (E-v p_z)}}{(e^{\beta \gamma (E-vp_z)}\pm1)^2}.
\end{eqnarray}

\subsection{Solutions to the hydrodynamical equations}

\subsubsection*{Detonations and deflagrations}

As mentioned, hydrodynamical treatments mostly model the plasma as a perfect ultrarelativistic fluid and solve the corresponding stress-energy conservation equations. When the bubble becomes large enough and the bubble wall propagates at a constant velocity in relation to the medium, the bubble profile becomes a \textit{similarity solution} to the equations, expanding linearly with time but maintaining its relative shape. The medium is assumed to be at rest both far ahead of and far behind the advancing bubble wall. This assumption dictates that no single, planar front may satisfy the stress-energy conservation equations (see e.g.~\cite{Megevand:2009gh}). Two possibilities are usually assumed to exist depending on whether the bubble wall advances at a velocity larger or smaller than the speed of sound in the medium $c_{s}$: The first case is generally known as a \textit{detonation}, the second as a \textit{deflagration}. In a detonation no information propagates ahead of the moving wall, which hits the medium while it is still at rest. The medium is heated up and accelerated by the bubble wall, then cools down and comes to rest in a \textit{rarefaction wave} while already in the broken symmetry phase behind the wall. In a deflagration, a \textit{shock front} propagates at or close to the speed of sound ahead of the bubble wall. It heats up and accelerates the fluid in the symmetric phase before the phase transition front proper brings it back to rest. The common assumption of modelling the phase transition front and the shock front (where it exists) as planar surfaces leads to constant values of the dynamic variables (fluid velocity and temperature) in the regions away from the fronts, including that between the phase transition front and the shock front in a deflagration. We will refine our calculation (for a deflagration) by splitting the bubble profile into three regions and writing the conservation equations separately for each. We adopt the planar approach to integrate the conservation equations across the bubble wall that separates the symmetric and broken phases, and across the shock front that propagates ahead of the wall. However (as in e.g.~\cite{Espinosa:2010hh}) we will take into account the sphericity of the bubble to calculate the evolution of the dynamic variables in the region between the fronts. In this way we may produce a whole profile of the bubble. We match our results to what we take to be the temperature of the universe outside the bubble (ahead of the shock front for a deflagration, or the bubble wall for a detonation). We carry out the hydrodynamical calculations for steady-state solutions. Finally, we investigate the possibility of runaway walls.

\subsubsection*{Hybrid solutions}

A more careful analysis (see e.g.~\cite{Espinosa:2010hh}) reveals the possibility of so-called \textit{hybrid} steady-state solutions in between detonations and deflagrations. Hybrid solutions have both a shock front propagating ahead of the phase transition front proper \textit{and} a rarefaction wave following the phase transition front. For a given physical situation they appear naturally with increasing steady-state wall velocity (arising with decreasing friction). A subsonic wall, as mentioned, propagates as a deflagration. If the wall becomes supersonic, however, it does not instantly become a detonation. Instead, a rarefaction wave appears behind the wall while a supersonic shock front still propagates ahead of the phase transition front. If the steady-state wall velocity becomes even larger, it eventually catches up with the shock front, which disappears, the solution becoming a pure detonation. In this work we do not concern ourselves with the dynamics of a rarefaction wave and therefore do not study in any detail the possibility of hybrid solutions. However, as noted below, our results are consistent with the appearance of hybrid solutions within the parameter space or our model.

\subsection{Dependence of the pressure on the wall velocity}

It has recently been established that the friction, which slows down the wall due to interactions between the wall and the particles in the plasma (and which exactly balances out with the pressure difference for a steady-state wall, see eq.~(\ref{eq:pressure}) ) reaches a \textit{finite limit} as $v_w \longrightarrow 1$ \cite{Bodeker:2009qy}. This opens up the possibility of runaway walls if the pressure difference is higher than this limit to the friction. In the ultrarelativistic limit both our 'slow wall' and 'fast wall' limits to the friction integral (equal to the steady-state pressure difference) converge to the same integral dependence as $E \approx p_z$ which tends indeed to a finite limit with $v_w$ approaching $1$. It must be noted, however, that the 'slow wall' limit of the friction integral contains the relaxation time $\tau$ which should not remain constant but grow as $v_w \longrightarrow 1$. Assuming a dependence $\tau \equiv \tau_0 \gamma$ with $\tau_0$ constant the friction in this limit diverges as $\gamma$. No such divergence occurs in the 'fast wall' limit. We assume therefore that our description is adequate in each limit. We will use this fact to link the relaxation time approximation with the hydrodynamical calculation of the wall velocity and produce a usable model for the friction in section 4.

\section{Hydrodynamical treatment}

\subsection{The conservation equations}

We essentially follow the same strategy as in \cite{Huber:2011aa}, originally adapted from \cite{Ignatius:1993qn}. We derive our hydrodynamical equations from the conservation of the energy-momentum tensor.

The energy-momentum tensor of the system consists of that of the electroweak plasma (modelled as a perfect relativistic fluid) plus that of the background Higgs field, 

\begin{eqnarray}
T^{\mu\nu}=T_{\rm{field}}^{\mu\nu}+T_{\rm{fluid}}^{\mu\nu}=\nonumber \\
=\partial^{\mu}\phi\partial^{\nu}\phi-g^{\mu\nu}\left(\frac{1}{2}\partial_{\alpha}\phi\partial^{\alpha}\phi\right)+\nonumber \\
+(\rho+P)u^{\mu}u^{\nu}-Pg^{\mu\nu}.
\end{eqnarray}

As the equation of state for the system we adopt $P\equiv P_{r}-V(\phi,T)$ with the radiation pressure $P_{r}\equiv aT^{4}=\frac{\pi^{2}}{90}g^{*}T^{4}$, $g^{*}$ being the number of effective degrees of freedom at the temperature $T$. We employ the usual thermodynamic relations

\begin{eqnarray}
P=-f\\
\rho=f-T\frac{\partial f}{\partial T},\label{Thermodynamic relations}
\end{eqnarray}
$f$ denoting the \textit{free energy per unit volume}. With these the total energy-momentum tensor may be expressed as

\begin{eqnarray}
T^{\mu\nu}=\partial^{\mu}\phi\partial^{\nu}\phi-g^{\mu\nu}\left(\frac{1}{2}\partial_{\alpha}\phi\partial^{\alpha}\phi\right)+\nonumber\\
+\left(\omega_{r}-T\frac{\partial V(\phi,T)}{\partial T}\right)u^{\mu}u^{\nu}-g^{\mu\nu}\left(P_{r}-V(\phi,T)\right)
\end{eqnarray}
where the \textit{radiative enthalpy} is $\omega_{r}\equiv4aT^{4}$. Even though this is conserved as a whole ($\partial_{\mu}T^{\mu\nu}\equiv0$),
the fluid and field components are not:

\[
\partial_{\mu}T^{\mu\nu}=\partial_{\mu}(T_{\rm{fluid}}^{\mu\nu}+T_{\rm{field}}^{\mu\nu})=0.\].

\begin{figure}
	\centering
		\includegraphics{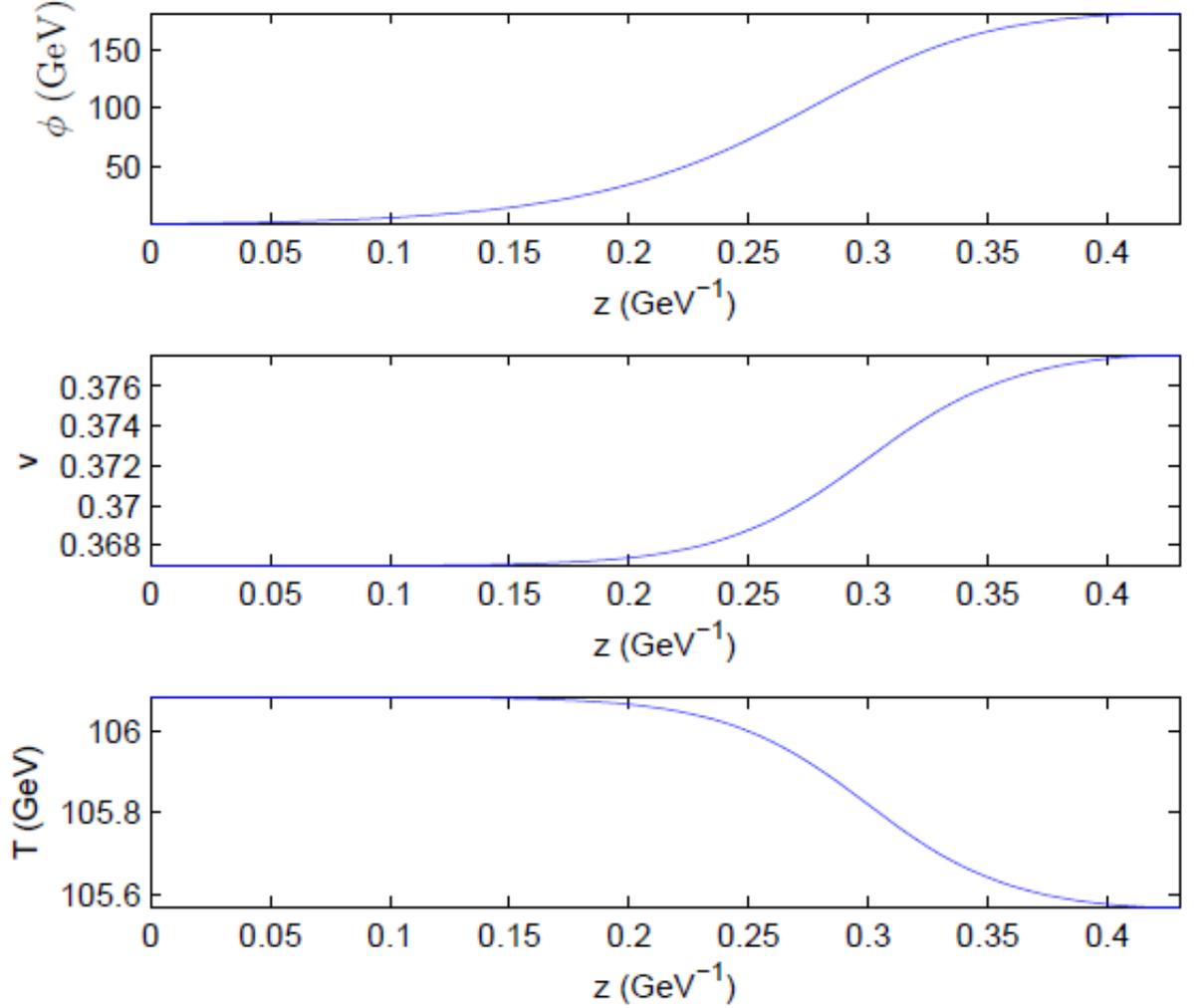}
	\caption{Higgs VEV $\phi$, velocity and temperature profiles across the (deflagration) bubble wall for the $\phi^6$ model with $M=800$ GeV, $m_h=115$ GeV at the temperature of the universe (ahead of the shock front) $T_u=105.49$ GeV and a value of the friction coefficient $\eta = 0.398$. The broken symmetry phase is on the right and the bubble wall propagates from right to left. The profile has been obtained solving the system (\ref{eqn:fluid1})-(\ref{eqn:fluid3}) via a linearisation procedure.}
	\label{fig:wall1}
\end{figure}

We now split the conservation equations into two parts (as in \cite{Ignatius:1993qn}), making each part equal to plus or minus an appropriately chosen friction term which we choose to make dependent on a dimensionless \textit{friction parameter}. Our original choice (later reconsidered, see \ref{subsec:frparandvw}) was inspired by the form of the equation of motion for the Higgs VEV in the 'slow wall' limit of the relaxation time approximation (\ref{eq:eomsw}), given the fermionic and bosonic mass dependence on the Higgs VEV. The conservation equations (in covariant form) may be split as

\begin{eqnarray}
\partial_{\mu}\left\{ \partial^{\mu}\phi\partial^{\nu}\phi-g^{\mu\nu}\left(\frac{1}{2}\partial_{\alpha}\phi\partial^{\alpha}\phi-V_{0}(\phi)\right)\right\} +\frac{\partial V_{1}(\phi,T)}{\partial\phi}\partial^{\nu}\phi\equiv-\eta \frac{\phi^2}{T_{\rm{s1}}}u^{\mu}\partial_{\mu}\phi\partial^{\nu}\phi\label{eq:splitting1}\\
\partial_{\mu}\left\{ \left(\omega_{r}-T\frac{\partial V_{1}(\phi,T)}{\partial T}\right)u^{\mu}u^{\nu}-g^{\mu\nu}P_{r}\right\} +\frac{\partial V_{1}(\phi,T)}{\partial T}\partial^{\nu}T\equiv+\eta \frac{\phi^2}{T_{\rm{s1}}}u^{\mu}\partial_{\mu}\phi\partial^{\nu}\phi\label{eq:splitting2}
\end{eqnarray}
where $V_{0}(\phi)$ is the part of the Higgs effective potential dependent only on the Higgs VEV while $V_{1}(\phi, T)$ is the part dependent also on temperature. $\eta$ is our friction parameter characterising the resistance of the plasma to the wall's movement and $T_{\rm{s1}}$ is a reference temperature, which we choose to be that of the plasma in the symmetric phase just ahead of the wall. To study the dynamics of the variables across the bubble and shock fronts we restrict ourselves to the $1+1$ dimensional case (the spatial dimension taken as perpendicular to the front, along the direction of bubble expansion). We assume steady-state solutions characterised by  constant bubble wall and shock front (where present) velocities along the spatial direction. Following \cite{Ignatius:1993qn} closely we arrive at the system (in the bubble wall's rest frame)

\begin{eqnarray}
\frac{d^2\phi(x)}{dx^2}=\frac{\partial V(\phi,T)}{\partial\phi}+\eta\frac{\phi^2}{T_{s1}}v\gamma\frac{d\phi(x)}{dx}\label{eqn:fluid1}\\
(4aT^4-T\frac{\partial V(\phi,T)}{\partial T})\gamma^2v=C_1\label{eqn:fluid2}\\
(4aT^4-T\frac{\partial V(\phi,T)}{\partial T})\gamma^{2}v^{2}+P_{r}-V(\phi,T)+\frac{1}{2}(\frac{d\phi}{dx})^{2}=C_2\label{eqn:fluid3}
\end{eqnarray}
where $v$ and $T$ in addition to $\phi$ are functions of the spatial coordinate and $\gamma$ is the usual relativistic factor $(1-v^2)^{-1/2}$. The value of the two arbitrary constants $C_1$ and $C_2$ is easiest to calculate in the symmetric phase, where $V$ as well as $\phi$ and its spatial derivatives vanish. We integrate equation (\ref{eqn:fluid1}) with boundary conditions

\begin{eqnarray}
	\phi '(x)=0 \text{  (at both ends of the integration interval)}\\
	\phi=0 \text{  (in the symmetric phase)}
\end{eqnarray}
at the same time solving for the velocity and temperature profiles across the bubble wall. We start from values of the velocity and temperature of the cosmic fluid in the symmetric phase ahead of the wall, $v_{\rm{s1}}$ and $T_{\rm{s1}}$, and reach values $v_{\rm{b}}$, $T_{\rm{b}}$ in the broken symmetry phase, behind the wall, see Fig \ref{fig:wall1}. These values are always expressed in the wall rest frame. These calculations can be done by a variety of methods e.g.~through a linearisation procedure described below.

\subsubsection*{The shock front}

The shock front (for a deflagration) propagates across the symmetric phase so both the Higgs VEV and the effective potential vanish. In the rest frame of the front the stress-energy conservation equations simplify to

\begin{gather}
\partial_x[(P+\rho)\gamma^2v]=0\nonumber\\
\partial_x[(P+\rho)\gamma^2v^2+P]=0\nonumber
\end{gather}
which integrate to

\begin{gather}
v_{\rm{u}}=\frac{1}{\sqrt{3}}\sqrt{\frac{3T_{\rm{s2}}^{4}+T_{\rm{u}}^{4}}{3T_{\rm{u}}^{4}+T_{\rm{s2}}^{4}}}\label{eq:sf1}\\
v_{\rm{s2}}=\frac{1}{3v_{\rm{u}}}\label{eq:sf2}
\end{gather}
The subscripts u, s2 refer to values in the symmetric phase respectively ahead of and behind the shock front.

\subsubsection*{The frame of the fluid}

For a subsonic, pure deflagration bubble\footnote{Note once again that we are not considering hybrid solutions.}, the velocities found in the rest frames of each front are related to the velocities in the rest frame of the fluid far away from the wall via the corresponding relativistic transformations:

\begin{gather}
v_{\rm{shock\;front}}=v_{\rm{\rm{u}}}\nonumber \\
v_{\rm{bubble\;wall}}=v_{\rm{\rm{b}}}\nonumber \\
v_{\rm{fluid}}=\frac{v_{\rm{b}}-v_{\rm{s1}}}{1-v_{\rm{b}}v_{\rm{s1}}}\label{eq:velocities}\end{gather}
$v_{\rm{fluid}}$ being the velocity of the fluid in the 'frame of the universe' (in which the fluid far ahead of and behind the bubble wall and shock front is at rest) just ahead of the bubble wall.

For a supersonic, pure detonation bubble which hits the medium while at rest (with no shock front) it is enough to integrate the system (\ref{eqn:fluid1})-(\ref{eqn:fluid3}) across the bubble wall, which gives us the wall velocity straight away:

\begin{align}
	v_{\rm{bubble\;wall}}=v_{\rm{s1}}.
\end{align}

\subsubsection*{Intermediate region in a deflagration bubble}

Taking the sphericity of a subsonic bubble into account requires integrating the conservation equations across the region between the bubble wall and the shock front. As we have seen, from the integration across the bubble wall we know the wall velocity. We now transform to the 'frame of the universe' and take advantage of the similarity properties of the expanding bubble to make all variables dependent on the single coordinate $\xi=r/t$. The stress-energy equations become \cite{Espinosa:2010hh}


\begin{eqnarray}
	\frac{\xi-v}{w}\frac{\partial \rho}{\partial \xi}=2\frac{v}{\xi}+\frac{\partial v}{\partial \xi}(1-\gamma^2(\xi-v))\nonumber\\
	\frac{1-v\xi}{w}\frac{\partial P}{\partial \xi}=\gamma^2(\xi-v)\frac{\partial v}{\partial \xi}.\nonumber
\end{eqnarray}
Isolating derivatives, one obtains
\begin{eqnarray}
\frac{dT}{d\xi}=\frac{2vT}{3\xi(1-v\xi)}[\frac{\xi-v}{1-v\xi}-\frac{1-v\xi}{3(\xi-v)}]^{-1}\label{eq:interm1}\\
\frac{dv}{d\xi}=\frac{2v(1-v^2)}{3\xi(\xi-v)}[\frac{\xi-v}{1-v\xi}-\frac{1-v\xi}{3(\xi-v)}]^{-1}.\label{eq:interm2}
\end{eqnarray}


We integrate along the radial direction, starting at the bubble wall and making an initial guess for $\xi$ at the shock front. Our guess is the shock front velocity so we can transform the final value of the fluid velocity $v_{\rm{s2}}$ (just behind the front) to the frame of the front and use (\ref{eq:sf2}) to find the fluid velocity ahead of the front. We take guesses until this matches $\xi$ for the front. We may finally calculate the temperature of the undisturbed universe $T_u$ via (\ref{eq:sf1}). It is worth mentioning that for a weak phase transition the 'leap' of the variables across the shock front becomes extremely small \cite{Moore:1995si}. In addition, because the calculation in the intermediate region is done in the 'frame of the universe' where the fluid velocity is very small (unless the phase transition is extremely strong) we may linearise and integrate (\ref{eq:interm1}) and (\ref{eq:interm2}) for small $v$, arriving at
\begin{eqnarray}
T(\xi)= T(\xi_0) e^{-\frac{2 v(\xi_0) \xi_0^2}{1-3\xi_0^2} (\frac{1}{\xi_0}-\frac{1}{\xi})}\label{eq:interm3}\\
v(\xi)= v(\xi_0) \left(\frac{\xi_0}{\xi}\right)^2 \left(\frac{3\xi^2-1}{3\xi_0^2-1}\right)\label{eq:interm4}
\end{eqnarray}
where $\xi_0 = v_{\rm{bubble\;wall}}$ and $v(\xi_0) = v_{\rm{fluid}}$. In this manner, provided we know $\eta$, it is possible to look for a self-consistent, steady-state solution to the hydrodynamical equations for which the temperature outside the bubble is whichever $T_u$ we wish to impose (the simplest assumption being to equate $T_u$ with the nucleation temperature $T_n$).

\begin{figure}
	\centering
		\includegraphics{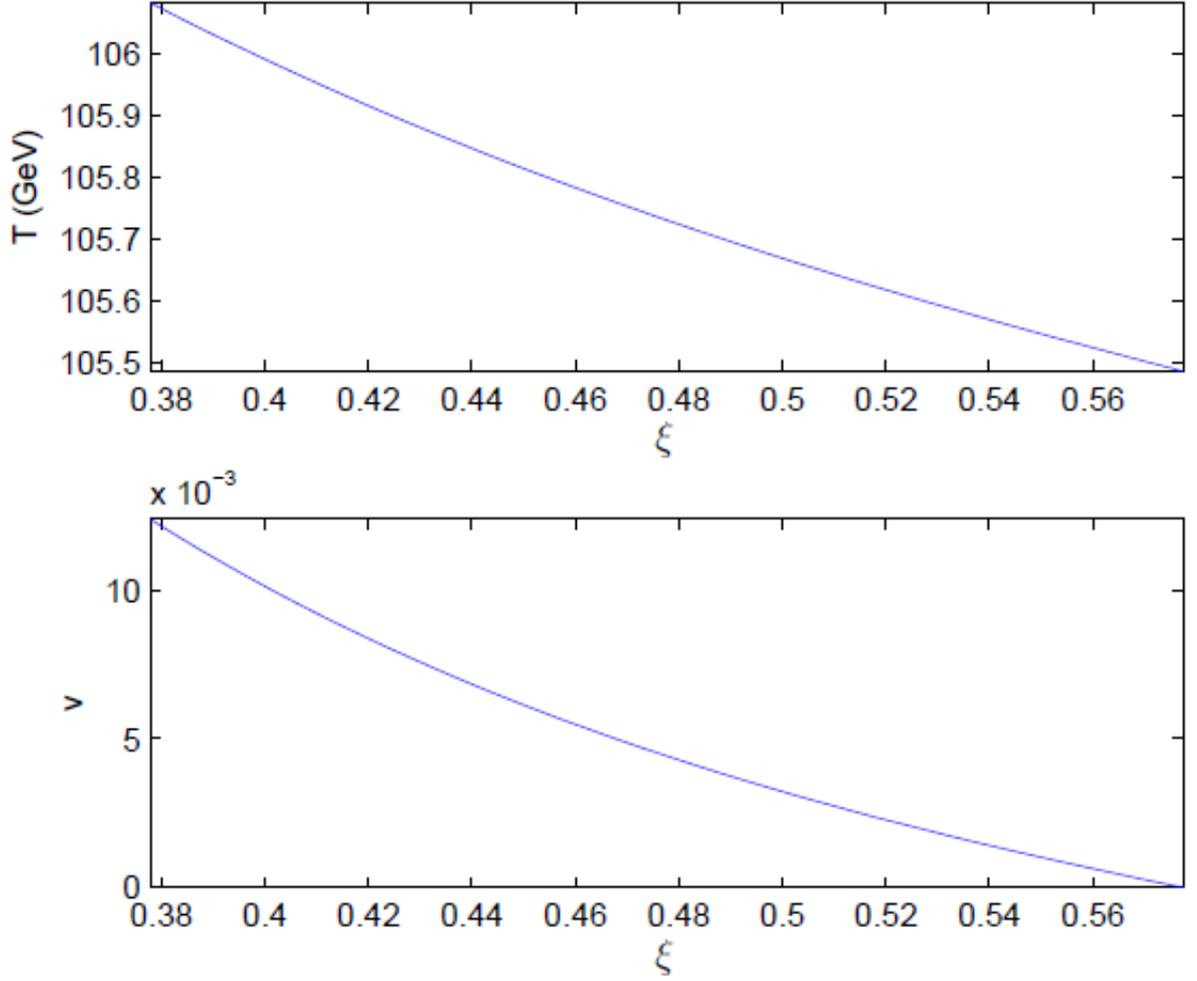}
	\caption{Velocity and temperature profiles across the intermediate region between the bubble wall and the shock front for the same deflagration bubble as in figure \ref{fig:wall1} with $M=800$ GeV, $m_h=115$ GeV at $T_u=105.49$ GeV. Here the position of the bubble wall is on the left end of the integration interval and that of the shock front on the right. If the sphericity of the bubble is neglected (and a less realistic planar approximation adopted instead), $v$ and $T$ do not vary across this region.}
	\label{fig:Tv_vs_xi}
\end{figure}

\subsection{The friction parameter and the wall velocity}\label{subsec:frparandvw}

In the above analysis $\eta$ is a Lorentz scalar which we take to be a number parametrising friction. This is what the covariant construction of the energy-momentum conservation equations suggests. However this leads to immediate trouble. Note that the same reasoning that took us to (\ref{eq:pressure}), if applied to (\ref{eqn:fluid1}), leads to a pressure difference \textit{divergent} as $v_w\longrightarrow1$ because of the $\gamma$ factor present. This may not be very important if the bubble wall is very subsonic (as it is in supersymmetric light stop scenario \cite{Carena:1996wj,Myint:1992wi,Espinosa:1993yi,Brignole:1993wv}) but for fast walls (or in order to investigate the possibility of runaway) this difficulty cannot be ignored. We are therefore forced to consider a more complex relationship between the friction parameter and the wall velocity. The simplest assumption that we can make is that the friction parameter depends on $v_w$ as $\eta(v_w) = \eta_0 \gamma^{-1}$, with $\eta_0$ constant and $\gamma = (1-v_w^2)^{-\frac{1}{2}}$, so that the $\gamma$ factor in (\ref{eqn:fluid1}) is effectively not there. This is the assumption we shall make in our calculations.

\section{Modelling the friction}

\subsection{Friction in the relaxation time approximation}

We wish to produce a reliable estimate for the friction parameter in (\ref{eqn:fluid1}) in order to solve for the bubble profile and calculate the wall velocity for our model. The form of the friction parameter supplied by the relaxation time approximation in its 'slow wall' limit can be written in analogy to (\ref{eqn:fluid1}) through the relevant mass couplings,
\begin{eqnarray}
	\frac{dm^2}{d\phi}\int\frac{d^3p}{(2\pi)^32E}\delta f(p,x)=
	\nonumber\\
	\frac{dm^2}{d\phi}\int\frac{d^3p}{(2\pi)^32E}\tau \beta \gamma v \frac{(m^2)'}{2E}\frac{e^{\beta \gamma (E-v p_z)}}{(e^{\beta \gamma (E-vp_z)}\pm1)^2}\sim
	\nonumber\\
	\phi^2 \phi' \tau \beta \gamma v \int\frac{d^3p}{(2\pi)^34E^2}\frac{e^{\beta \gamma (E-v p_z)}}{(e^{\beta \gamma (E-vp_z)}\pm1)^2},\label{eq:reltfrterm1}
\end{eqnarray}
for one degree of freedom. In the last step we assumed that $m$ is proportional to $\phi$. The relevant prefactors coming from $\tau$ and the mass couplings depend on the model considered. Rather than carrying out a first-principles calculation, our {\em goal} is to develop a technique to relate the existing results from \cite{Moore:1995si} (arrived at using the full form of the collision integral) to any Standard Model-like situation (in which  friction is carried by the top and weak gauge boson degrees of freedom). A similar analysis for the case of friction dominated by light stops was carried out in \cite{Huber:2011aa}.

Note that we are free to change integration variables in (\ref{eq:reltfrterm1}) from $p \rightarrow p'=\frac{p}{T}$. Since $E = \sqrt{m^2+p^2}$ the friction term suggested by the relaxation time approximation in this form becomes
\begin{eqnarray}\label{eq:mominteta}
	\frac{dm^2}{d\phi}\int\frac{d^3p}{(2\pi)^32E}\delta f(p,x)\sim
	\nonumber\\
	\phi^2 \phi' \tau \gamma v \int\frac{d^3p'}{(2\pi)^3 \left((\frac{m}{T})^2+p'^2\right)}\frac{e^{\gamma (\sqrt{(\frac{m}{T})^2+p'^2}-\frac{v p_z}{T})}}{(e^{\gamma (\sqrt{(\frac{m}{T})^2+p'^2}-\frac{v p_z}{T})}\pm1)^2}.
	\nonumber\\
\end{eqnarray}

\begin{figure}[htp]
  \begin{center}
  \includegraphics[scale=0.5]{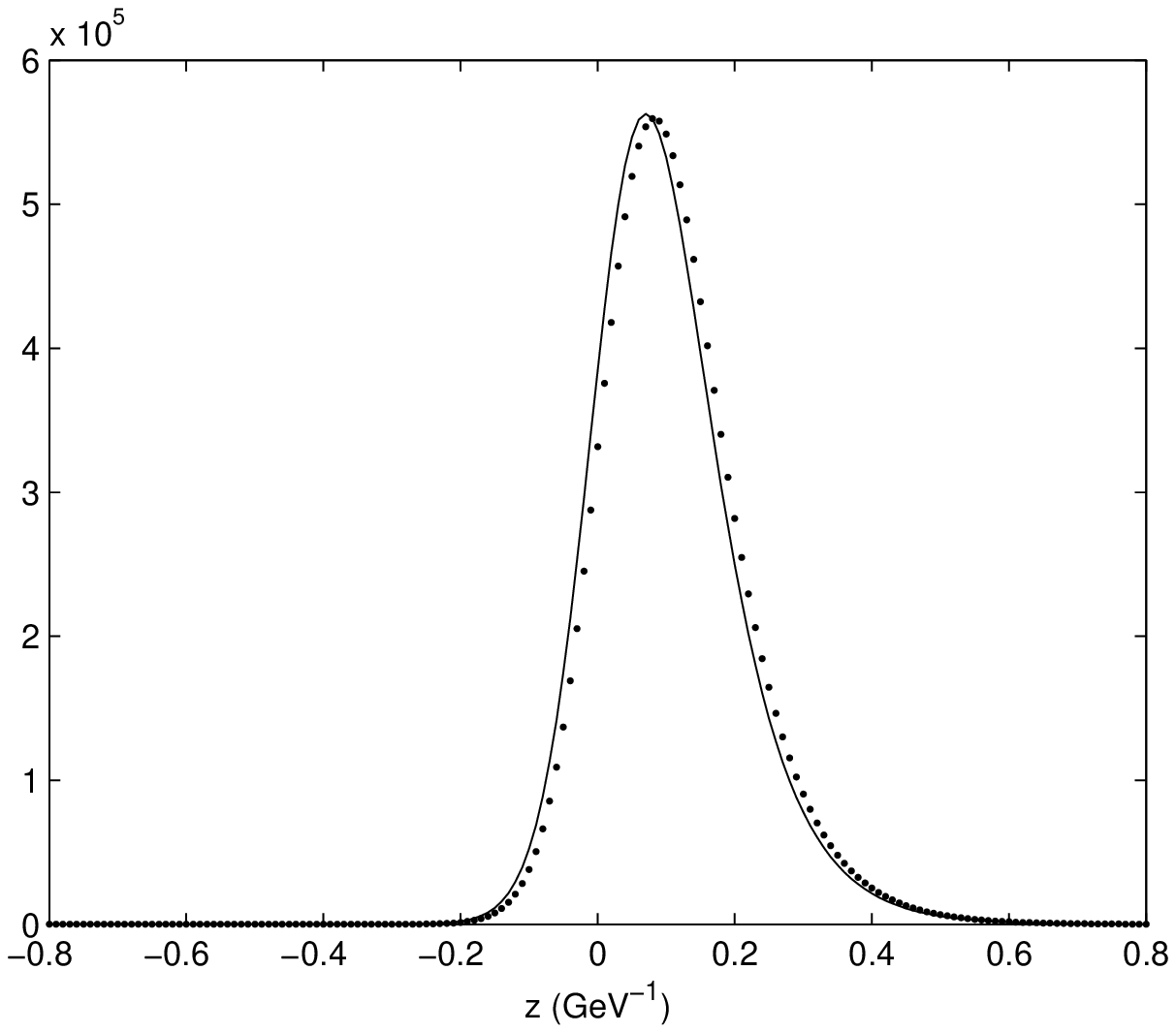}
    \includegraphics[scale=0.5]{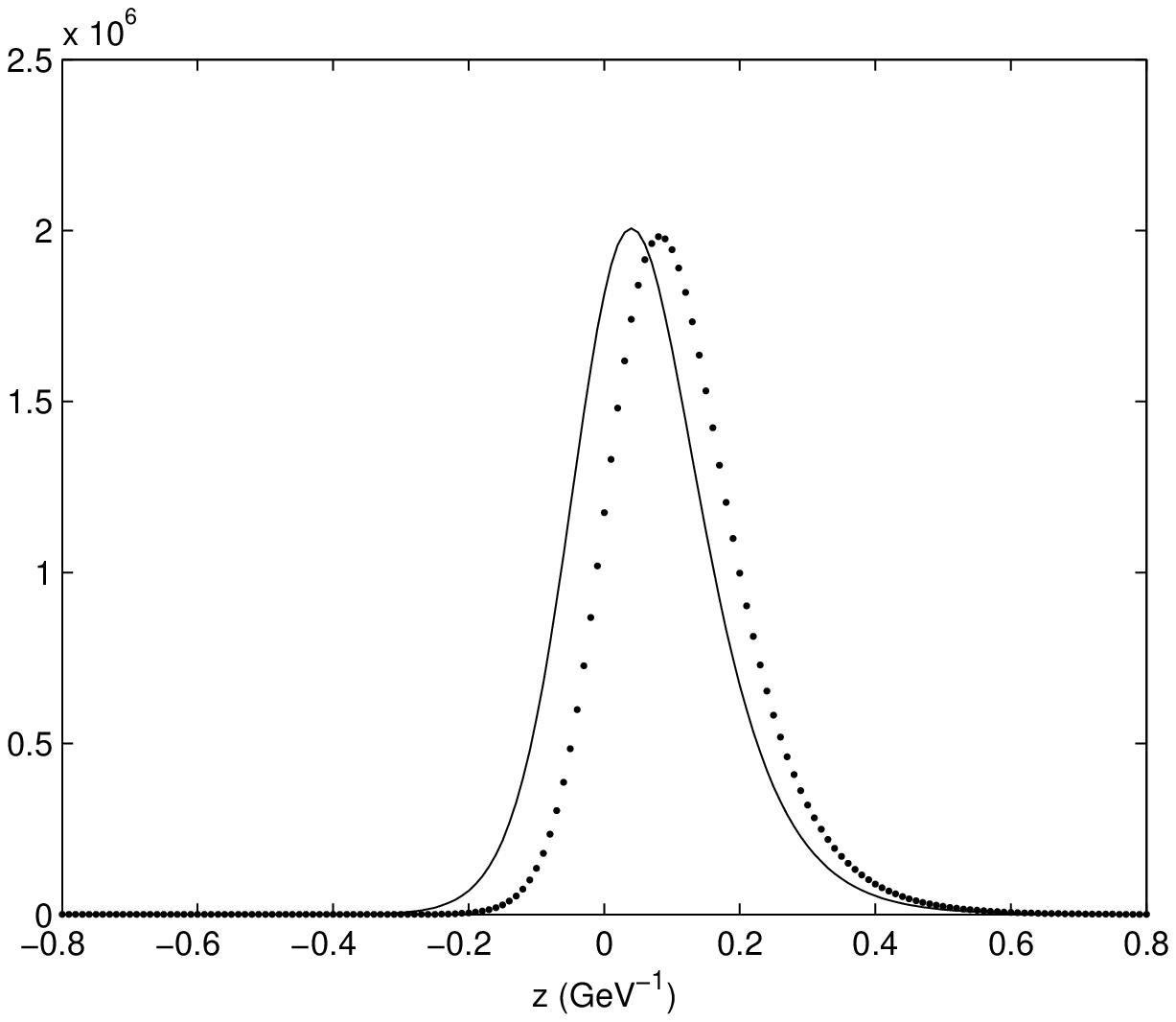}
  \end{center}
  \caption{Spatially varying part of the friction term according to the relaxation time approximation (slow wall limit),  $\phi^2 \phi' \int\frac{d^3p}{(2\pi)^3 (2E)^2} \frac{e^{\beta \gamma (E-v_w p_z)}}{(e^{\beta \gamma (E-v_w p_z)}\pm1)^2}$ (solid line), and the same term with the momentum integral replaced by a fitted constant (dotted line), $C\;\phi^2 \phi'$, for fermions (left) and bosons (right) in an example case. We use the hyperbolic tangent Ansatz to approximate the bubble profile and assume $\phi_0 = 100$ GeV in the broken symmetry phase, $T = 100$ GeV, $L_w \cdot T = 15$, and a mass dependence $m \equiv \frac{1}{\sqrt{2}} \phi$.}
  \label{fig:frictermfitted}
\end{figure}

\begin{figure}
  \begin{center}
 \includegraphics{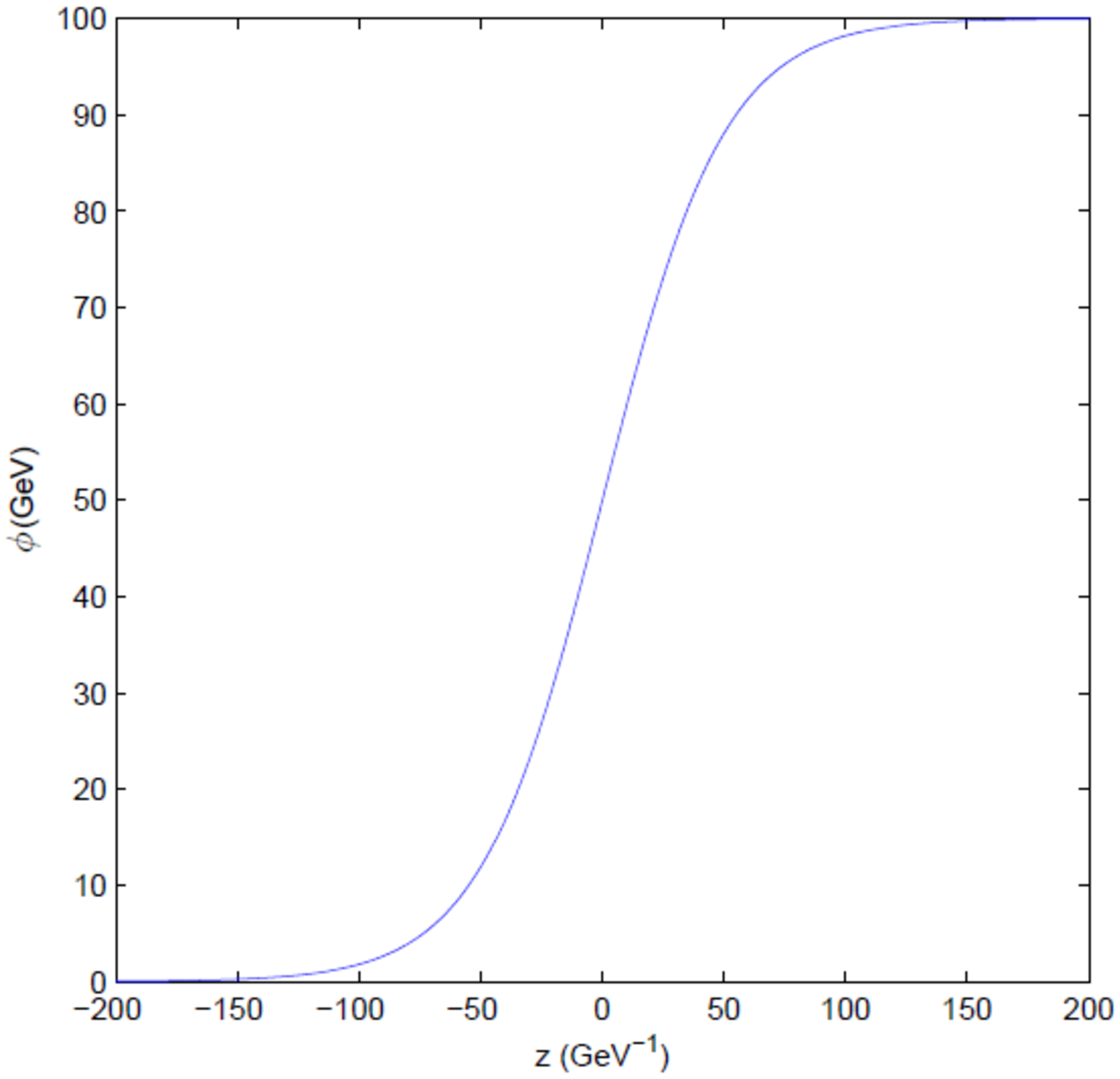}
 \caption{Sample hyperbolic tangent Ansatz for the Higgs VEV across the bubble wall. The broken symmetry phase is on the right hand side with an assumed VEV of $\phi_0 = 100$ GeV. We take $L = 50$ GeV$^{-1}$.}
 \label{fig:tanhAnsatz}
 \end{center}
\end{figure}

Given the general dependence of the mass the integral contains factors of $\frac{\phi}{T}$. This strongly suggests that friction is dependent on the strength of the phase transition, commonly expressed in the literature by the parameter $\xi_c = \frac{v_c}{T_c}$, that is, $\frac{\phi_0}{T}$ evaluated at the critical temperature. If we wish to approximate the friction integral by a single numerical contribution (from each degree of freedom), one way would be to approximate the  $\frac{\phi}{T}$ factors by $\frac{v_c}{T_c}$ (or an appropriate related value like $\frac{1}{2}$ $\frac{v_c}{T_c}$, the value of $\frac{\phi}{T}$ roughly half way across the wall at $T_c$) and integrate over momentum\footnote{Note that there are other choices as regards expressing the strength of the phase transition, namely as a function of the nucleation ($\xi_n = \frac{v_n}{T_n}$) or broken symmetry phase ($\xi_b = \frac{v_b}{T_b}$) temperatures.}. Alternatively, one can plot the spatial dependence of the friction term itself (peaked because of the $\phi'$ factor, which goes to zero away from the bubble wall on either side) and replace the integral by a constant which equalises the peaks (Figure \ref{fig:frictermfitted}). This can be done, for example, through the 2-parameter hyperbolic tangent Ansatz commonly used to approximate the shape of the wall (see e.g.~\cite{Moore:1995si}). The variation of the Higgs VEV across the bubble wall (assumed planar) is given in this approximation by (Fig \ref{fig:tanhAnsatz})
\begin{eqnarray}
\phi(z) = \frac{\phi_0}{2}\left(1 + \tanh \frac{z}{L}\right)
\end{eqnarray}
where $L$ is the wall thickness and $\phi_0$ the value of the Higgs VEV in the broken
symmetry phase behind the bubble wall as read from the effective potential. As the Ansatz is written in the rest frame of the advancing steady-state bubble wall there is no time dependence. A quick way to obtain a usable value for $L$ is through
\begin{eqnarray}
L^2 = \frac{\phi_0^2}{8 V_b}
\end{eqnarray}
with $V_b$ the height of the potential barrier between the two minima of the effective potential (this is suggested by the simplified 2-parameter scalar potential $V(\phi) = \frac{\lambda}{4} \phi^2 (\phi - \phi_0)^2$ with the equation of motion $\frac{d^2 \phi}{dz^2} = \frac{\partial V(\phi)}{\partial \phi}$, into which the hyperbolic tangent Ansatz may be substituted).

Using either of these approaches, or a comparable simplification, the relaxation time approximation model supplies a numerical coefficient for a description of the friction of the form $\eta(v_w) \equiv \eta_0 \gamma^{-1}$. Because we assume that $\eta_0$ has no dependence on $v_w$ we set $v_w \equiv 0$ in the momentum integral in (\ref{eq:mominteta}).

\subsection{Fixing the friction parameter}

As mentioned, the prefactors for the friction integral for each massive degree of freedom in the plasma that couples strongly to the Higgs can be calibrated from existing results, in our case those for the Standard Model in \cite{Moore:1995si}. This is because  friction processes are a function of the particle content of the model and the relevant particle interactions, and therefore, as long as we remain within a Standard Model-like situation, the prefactors to the momentum integrals (the relevant coupling constants and the chosen form for the relaxation time approximation) should hold. This is, for example, the case in the dimension-6 extension of the SM studied below. 

The standard model 2-loop, high-temperature thermal effective potential used in \cite{Moore:1995si} is
\begin{equation}
V_{\rm{eff}}(\phi,T)=D(T^2-T_0^2)\phi^2-CT^2\phi^2\ln(\frac{\phi}{T})-ET\phi^3+\frac{\lambda}{4}\phi^4
\end{equation}
with
\begin{eqnarray}
	\lambda&=&\frac{m_h^2}{2v_0^2}-\frac{3}{16\pi v_0^4}[2m_w^4\ln(\frac{m_w^2}{a_bT^2})+m_z^4\ln(\frac{m_z^2}{a_bT^2})-4m_t^4\ln(\frac{m_t^2}{a_fT^2})]\nonumber\\
	D&=&\frac{1}{8v_0^2}(2m_w^2+m_z^2+2m_t^2)\nonumber\\
	C&=&\frac{1}{16\pi^2}(1.42g_w^4+4.8g_w^2\lambda-6\lambda^2)\nonumber\\
	E&=&\frac{1}{12\pi}[4(\frac{m_w}{v_0})^3+2(\frac{m_z}{v_0})^3+(3+3^{1.5})\lambda^{1.5}]\nonumber\\
	B&=&\frac{3}{64 \pi^2 v_0^4}(2m_w^4+m_z^4-4m_t^4)\nonumber\\
	T_0&=&\sqrt{\frac{1}{4D}(m_h^2-8Bv_0^2)}\nonumber
\end{eqnarray}
and
\begin{eqnarray}
	g_w&=&\frac{2m_w}{v_0}\nonumber\\
	m_w&=&80.4 \rm{\ GeV} \nonumber\\
	m_z&=&91.2 \rm{\ GeV}\nonumber\\
	m_t&=&174.0 \rm{\ GeV}\nonumber\\
	v_0&=&246.0 \rm{\ GeV}\nonumber\\
	a_b&=&49.78019250\nonumber\\
	a_f&=&3.111262032.\nonumber\\
\end{eqnarray}

We show the values of $\eta_0$ which reproduce the values of $v_w$ in \cite{Moore:1995si} in table \ref{table:fitetaphi6} (denoted by $\eta_{0,SM}$). We now wish, as mentioned, to produce a prediction for the friction coefficient for our model. The authors of \cite{Moore:1995si} estimate $60 \%$ of the Standard Model friction to come from fermions and $40 \%$ from bosons. We call the values of the bosonic and fermionic numerical coefficients obtained from the momentum integrals in eq (\ref{eq:mominteta}) for each case in \cite{Moore:1995si} (and characterised by a value of the strength of the phase transition $\xi = \frac{\phi_0}{T}$) $I_{0b}(\xi)$, $I_{0f}(\xi)$. With this calibration, the friction parameter for SM-like friction, e.g.~within the dimension-6 model (characterised by a specific value of the strength of the phase transition $\xi$) may be found as
\begin{eqnarray}
	\eta_0(\xi) = \eta_{0,SM} \left(0.6 \frac{I_f (\xi) }{I_{0f} (\xi_0) } + 0.4 \frac{I_b (\xi) }{I_{0b} (\xi_0) } \right).\label{eq:etaphi6}
\end{eqnarray}
\begin{table}
\caption{Values of the Standard Model quartic coupling $\lambda$, Higgs mass, strength of the phase transition $\xi_n = \frac{v_n}{T_n}$, nucleation temperature and fitted friction coefficient $\eta$ for relevant cases in \cite{Moore:1995si}.}
\centering
\bigskip
\begin{tabular}{c c c c c c}
\hline\hline
$\lambda_T$ & $m_h \rm{ (GeV)}$ & $\xi_n = \frac{v_n}{T_n}$ & $v_w \rm{ (from} \cite{Moore:1995si})$ & $T_n \rm{ (GeV)}$ & $\eta_{0,SM} \rm{ (fitted)}$\\
\hline
0.023 &  0 & 0.98 & 0.374 &  57.192 & 0.5628 \\
0.030 & 50 & 0.80 & 0.392 &  83.426 & 0.5286 \\
0.041 & 68 & 0.65 & 0.412 & 100.352 & 0.6431 \\
0.050 & 79 & 0.58 & 0.428 & 111.480 & 0.6705 \\
0.060 & 88 & 0.53 & 0.441 & 120.934 & 0.6707 \\
\end{tabular}
\label{table:fitetaphi6}
\end{table}

\section{The dimension-6 extension to the SM}

It has been suggested \cite{Zhang:1992fs,Dine:1990fj,Zhang:1994fb} that new physics may appear as higher-dimensional, nonrenormalisable operators added to the Standard Model Higgs scalar potential, getting around the constraints posed by present experimental bounds on the Higgs mass and providing additional sources of CP violation. The addition of dimension-6 operators to the Higgs potential \cite{Bodeker:2004ws,Grojean:2004xa,Ham:2004zs} has been proposed, numerical calculations having shown that further higher-order terms suppressed by the same low cut-off scale as the dimension-6 terms give corrections of only a few percent to the strength of the phase transition $\frac{v_c}{T_c}$ \cite{Grojean:2004xa}. The dynamics of the electroweak phase transition in this scenario are parametrised by the Higgs boson mass $m_H$ and the cut-off scale $M$. In this situation the quartic coupling of the Higgs potential may assume negative values. Such dimension-6 operators could stem from integrating out a massive degree of freedom like a scalar singlet \cite{Grojean:2004xa}, or alternatively from sources such as strongly coupled gravity \cite{Bodeker:2004ws}. New physics with a comparatively low cut-off scale may lead to non-standard signals which could be detected in the near future, such as modified Higgs self-couplings \cite{Grojean:2004xa}.

\subsection{The effective potential}

As dimension-6 effective potential we take \cite{Bodeker:2004ws}
\begin{eqnarray}
V_{\rm{eff}}(\phi,T)=\frac{1}{2}[-\mu^2+(\frac{1}{2}\lambda+\frac{3}{16}g_2^2+\frac{1}{16}g_1^2+\frac{1}{4}y_t^2)T^2]\phi^2\nonumber\\
-\frac{g_2^3}{16\pi}T\phi^3+\frac{\lambda}{4}\phi^4+\frac{3}{64\pi^2}y_t^4\phi^4\ln\left(\frac{Q^2}{c_FT^2}\right)\nonumber\\
+\frac{1}{8M^2}(\phi^6+2\phi^4T^2+\phi^2T^4)
\end{eqnarray}
where $Q\equiv m_{\rm{top}}$ and $c_F\approx13.94$. $M$ and $m_h$ are the free parameters of the model. $\mu$ and $\lambda$ are found through the conditions for the zero-temperature potential ($v_0=246$ GeV)
\begin{eqnarray}
	V_{\rm{eff}}(\phi,0)=-\frac{\mu^2}{2}\phi^2+\frac{\lambda}{4}\phi^4+\frac{1}{8M^2}\phi^6-\frac{3}{64\pi^2}y_t^4\phi^4\ln\left(\frac{y_t^2\phi^2}{2Q^2}\right)
\end{eqnarray}
which take the form
\begin{eqnarray}
\left.\frac{\partial V_{\rm{eff}}(\phi,0)}{\partial\phi}\right|_{\phi=v_0}=0 \text{,\ \ \ \ \ \ } \left.\frac{\partial^2V_{\rm{eff}}(\phi,0)}{\partial\phi^2}\right|_{\phi=v_0}=m_h^2.
\end{eqnarray}

\section{Results for the dimension-6 extension}

\subsection{General solutions to the hydrodynamic equations.}

\begin{figure}
	\centering
		\includegraphics{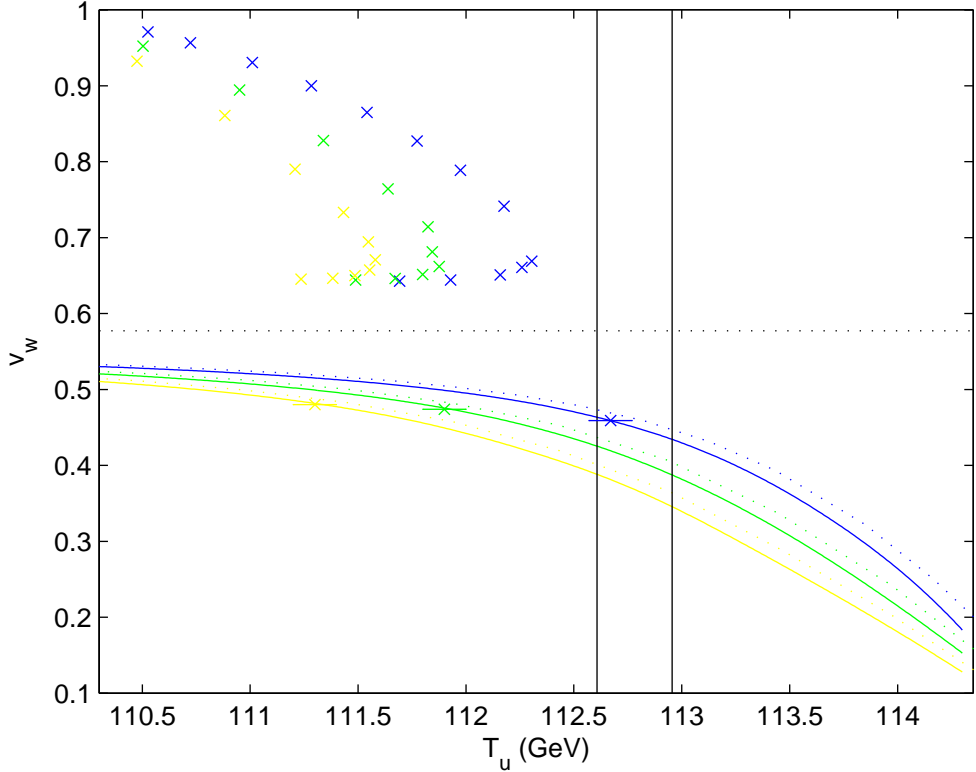}
	\caption{Steady-state bubble wall velocity vs temperature of the universe for three values of the friction coefficient $\eta$, 0.3 (blue), 0,4 (green) and 0.5 (yellow) for the dimension-6 model with $M=800$ GeV, $m_h=120$ GeV. The continuous lines below the horizontal line which marks the speed of sound are subsonic solutions, the dotted lines marking solutions found through neglecting the sphericity of the bubble. At the horizontal error bars the stability criterium for subsonic solutions changes sign (the stability region lies to the right of the mark). The crosses above the horizontal lines indicate supersonic solutions in which the bubble wall hits the medium at rest. The two vertical lines mark the nucleation and finalisation temperatures for the phase transition for these parameters. Note the two branches of supersonic solutions for each $\eta$. In this example and for these values of $\eta$ supersonic steady-state solutions would be excluded but stable subsonic ones allowed throughout the duration of the phase transition.}
	\label{fig:v_vs_T}
\end{figure}

We begin by studying the dependence of the steady-state wall velocity for pure deflagration and detonations solutions (excluding hybrids) on arbitrary values of the friction parameter for a specific choice of dimension-6 model parameters. We present the results of such an analysis for an arbitrary choice of model parameters in Figure \ref{fig:v_vs_T}, expressing the wall velocity as a function of the temperature of the universe $T_u$ for values of $\eta = 0.3$, $0.4$, $0.5$. The general shape of our solutions agrees with previous studies (see e.g.~\cite{Megevand:2009ut}). At high values of the temperature of the universe only subsonic solutions are allowed. The wall velocity increases with decreasing $T_u$ and is obviously higher for lower values of $\eta$, becoming \textit{larger} if we choose to neglect the sphericity of the expanding bubble and assume planar symmetry in the region between the shock front and the bubble wall\footnote{As noted, with such a simplification the dynamic variables $v$, $T$ do not change across this region.}. As $T_u$ keeps decreasing two branches of additional, supersonic solutions appear. Of these only the upper branch is physical (note that the shape of the lower branches implies that $v_w$ would \textit{decrease} with lower $\eta$, and again decrease as $T_u$ decreases and the free energy difference released by the phase transition becomes larger).

\begin{figure}
	\centering
		\includegraphics{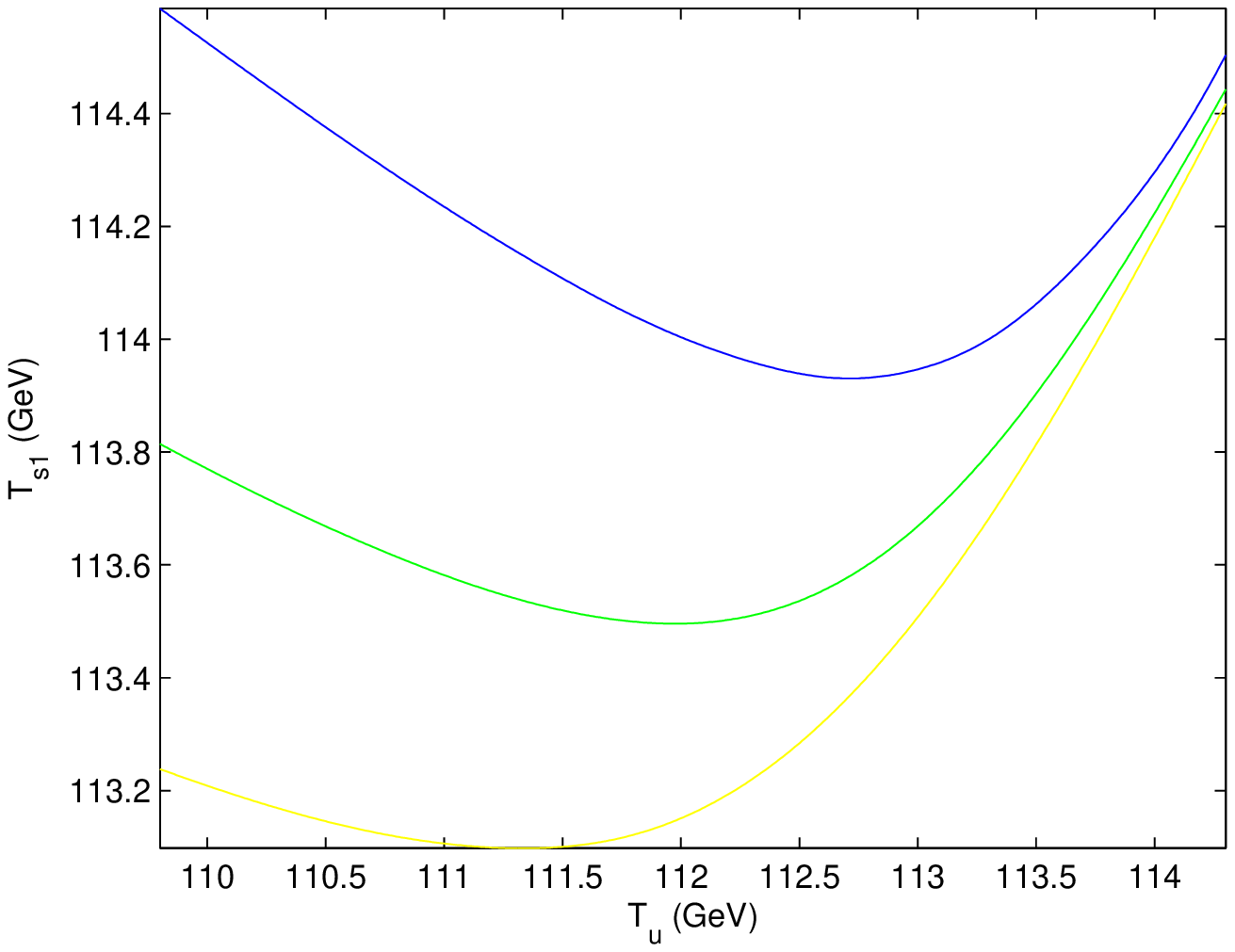}
	\caption{Temperatures in the symmetric phase just ahead of the bubble wall vs plasma temperature for the parameters of figure (\ref{fig:v_vs_T}) and values of the friction parameter $\eta=0.3, 0.4, 0.5$ (blue, green and yellow). Right of the minimum $T_{s1}$ for each $\eta$ is the stability region for subsonic solutions according to the criterion in \cite{Huet:1992ex}.}
	\label{fig:Tq_vs_T}
\end{figure}

A related issue is the stability of the expanding bubbles, examined in \cite{Huet:1992ex}. That reference concludes that subsonic bubbles may become unstable to the appearance of perturbations larger than a critical size provided that, essentially, the wall velocity decreases with decreasing $T_{s1}$. We may equivalently plot $T_u$ vs $T_{s1}$ (Figure \ref{fig:Tq_vs_T}), finding the values of $T_u$ for each value of the friction coefficient below which subsonic bubbles may become unstable. For the choice of model parameters in Figure \ref{fig:v_vs_T}, and as a general rule, this is roughly the point at which supersonic solutions first appear, and also the region in which hybrid solutions would naturally exist. Low-velocity subsonic solutions for high $T_u$ are always stable.

\subsection{The wall velocity in realistic cases.}

We conclude by producing predictions for the numerical value of the friction parameter in a realistic setting and, through our full hydrodynamic calculation, for the wall velocity. Relying on recent LHC results we do this for a fixed Higgs mass $m_H = 125$ GeV and study the variation of the friction and $v_w$ with the cut-off scale $M$. We wish to investigate the impact on our studies of the criterion for runaway walls from \cite{Bodeker:2009qy}, which establishes that, if the slowing effect of the plasma on the advancing wall in the limit $v_w \rightarrow 1$ is not sufficient to counter the accelerating force of the vacuum potential, the wall will run away. The phase transition becomes stronger, and the bubble wall faster, with decreasing $M$. Correspondingly we find that (for a Higgs mass of $125$ GeV) the runaway criterion is satisfied for $M \lesssim 582$ GeV and a strength of the phase transition $\xi_n = \frac{v_n}{T_n} \gtrsim 3.3$.

\begin{figure}
	\centering
		\includegraphics{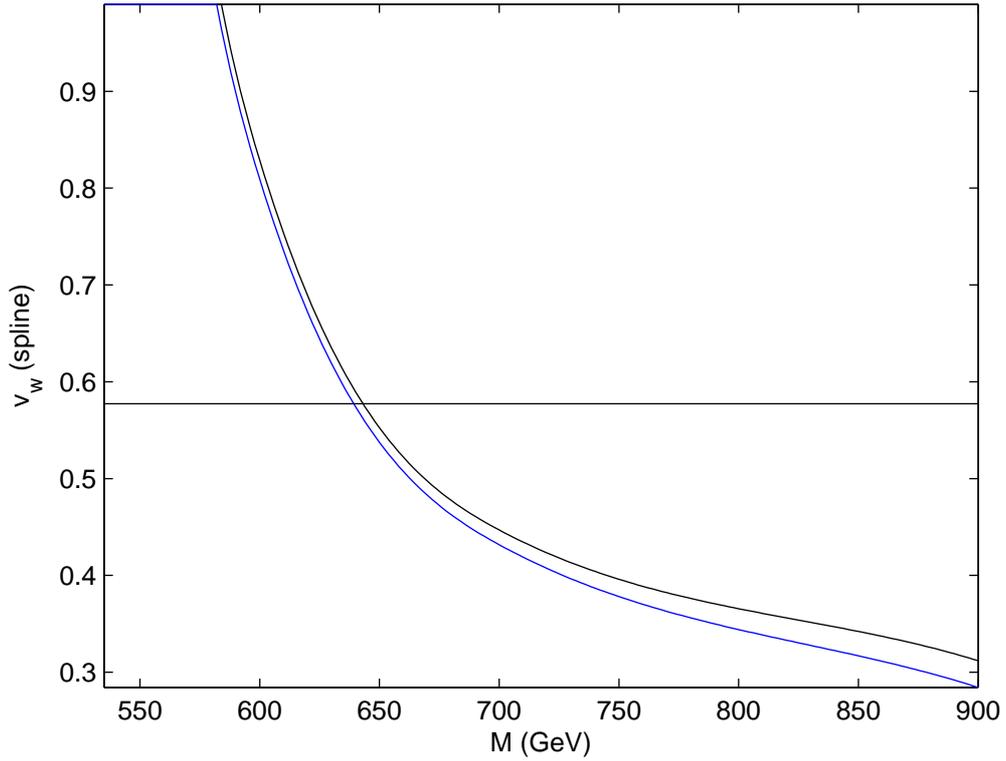}
	\caption{Wall velocity in the dimension-6 model as a function of the cutoff scale $M$ for $m_H = 125$ GeV calculated through the friction parameter obtained from the Standard Model calibration (in black, above) and the alternative calibration from the runaway criterion in \cite{Bodeker:2009qy} (in blue, below). The horizontal line marks the speed of sound in the medium. Wall velocities predicted in this way become supersonic below approximately $M \approx 640$ GeV. Note, however, that our linearized treatment of the hydrodynamic equations breaks down close to the speed of sound.}
	\label{fig:v_vs_M_125_sp}
\end{figure}

\begin{figure}
	\centering
		\includegraphics{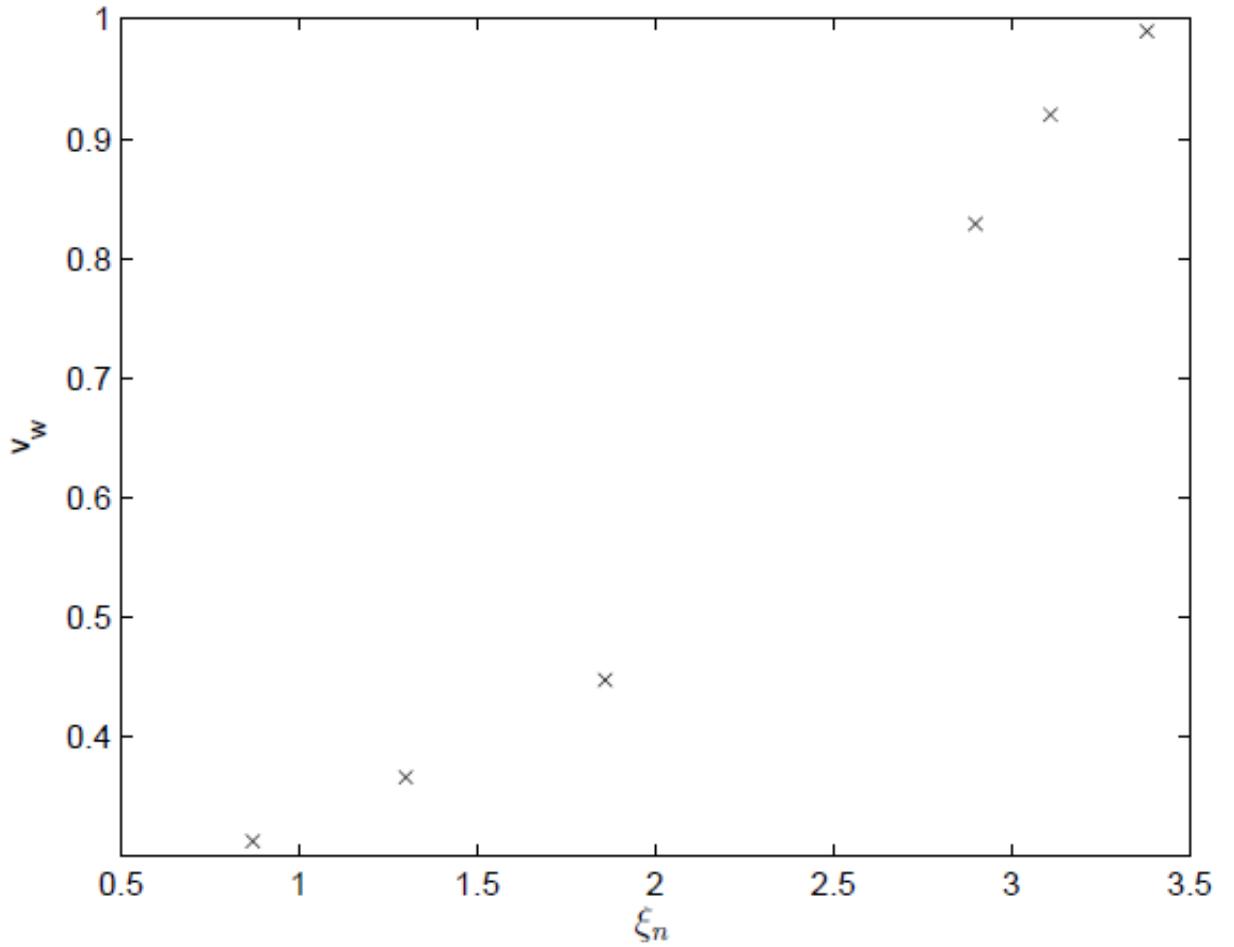}
	\caption{Wall velocity in the dimension-6 model for $m_H = 125$ GeV as a function of the strength of the phase transition at the nucleation temperature $\xi_n=\frac{\phi_0}{T_n}$.}
	\label{fig:v_vs_xin_125}
\end{figure}

The runaway criterion also provides us with an \textit{alternative} calibration point for the friction parameter. Instead of relying on values of the parameter fitted to existing Standard Model wall velocity calculations (as we wrote in eq.~(\ref{eq:etaphi6})) we may calculate the value of the friction parameter $\eta_{\rm{runaway}}$ that gives us $v_w \rightarrow 1$ for $M \sim 582$ GeV, $m_H = 125$ GeV, and write the calibration formula based on that $\eta_{\rm{runaway}}$ and the corresponding value of the strength of the phase transition. In Table \ref{table:results} we present the values of the friction parameter calculated on the basis of both calibrations (from existing Standard Model results and from our runaway point), which show remarkable consistency. We provide the values of the wall velocity calculated on the basis of our original calibration (based on the results in \cite{Moore:1995si}) for a range of $M$ values. The same results for the wall velocity are plotted in figure \ref{fig:v_vs_M_125_sp} alongside, for comparison, approximate values for $v_w$ found using the corresponding friction parameter values from the runaway point calibration (These approximate velocities are found by taking advantage of the fact that calibrated friction parameter values and wall velocities show a rough correspondence $\eta \cdot v_w \approx \rm{constant}$). Figure \ref{fig:v_vs_xin_125} shows the calculated wall velocities as a function of the strength of the phase transition at the nucleation temperature $\xi_n$. We want to stress that our linearized treatment of the hydrodynamic equations breaks down close to the speed of sound. As a result no cases in Figure \ref{fig:v_vs_xin_125} have wall velocities close to the speed of sound.

\begin{table}
\caption{Values of the friction coefficient and the wall velocity for the dimension-6 extension to the Standard Model, as a function of the cut-off scale $M$. Two sets of values of the friction coefficient are given, both calculated through the relaxation time approximation based on: 1) Existing Standard Model wall velocity calculations ($\eta_{\rm{SM}}$), and 2) The runaway criterion in \cite{Bodeker:2009qy} ($\eta_{\rm{r}}$). The wall velocity is calculated through the Standard Model-based value of the friction parameter, $\eta_{\rm{SM}}$. An approximate value of $v_w$ for $\eta_{\rm{runaway}}$ is also shown. We assume a Higgs mass $m_h = 125$ GeV. For each value of $M$ the nucleation temperature and the corresponding strength of the phase transition $\xi_n=\frac{\phi_0}{T_n}$ are given.}
\centering
\bigskip
\begin{tabular}{c c c c c c c}
\hline\hline
$M$ & $T_n$ & $\xi=\phi_0/T_n$ & $\eta_{\rm{SM}}$ & $\eta_{\rm{r}}$ & $v_w(\eta_{\rm{SM}})$ & $v_w(\eta_{\rm{r}})$\\
\hline
900 & 130.08 & 0.87 & 0.615 & 0.674 & 0.312 & 0.284\\
800 & 120.57 & 1.30 & 0.458 & 0.487 & 0.366 & 0.344\\
700 & 106.28 & 1.86 & 0.340 & 0.352 & 0.447 & 0.432\\
600 &  80.06 & 2.90 & 0.203 & 0.208 & 0.829 & 0.809\\
590 &  75.59 & 3.11 & 0.184 & 0.189 & 0.920 & 0.898\\
\end{tabular}
\label{table:results}
\end{table}

\section{Conclusions}
The bubble expansion velocity is a crucial ingredient in any computation of relic signals from a first-order electroweak phase transition. For instance, standard electroweak baryogenesis relies on a subsonic wall to allow for efficient diffusion of chiral charges into the symmetric phase\footnote{See, however, ref.~\cite{Caprini:2011uz} for an interesting alternative,}. Strong gravitational waves signals, on the other hand, require a supersonic bubble wall.

The microscopic determination of the wall velocity, even in the semiclassical approximation, is a very difficult task, which requires the simultaneous solution of the Higgs equation of motion and Boltzmann equations describing the out of equilibrium plasma. To date this has been achieved only for the cases of the SM \cite{Moore:1995si} and the light stop scenario of the MSSM \cite{John:2000zq}.

The value of the approach developed in this work is that it uses the results of these microscopic computations and allows to transfer them to related scenarios, where the Higgs dynamics is different but friction remains essentially the same as in a)  \cite{Moore:1995si} (dominated by tops and weak gauge bosons) or b) \cite{John:2000zq} (dominated by light stops). Other prominent extensions of the SM, such as the Two-Higgs-doublet model or SUSY models without light stops should, for instance, be quite well described by case a).

The {\em computation of the wall velocity}, as laid out in this work, consists of series of steps: Firstly, the temperature of the phase transition has to be determined as described in section 2.2. The main equations to consistently solve then are (34) -- (36), which describe the Higgs field coupled to an ideal fluid characterised by a fluid velocity and temperature. The unknown friction parameter $\eta$ can be obtained from eq.~(53), where the momentum integrals $I_{f,b}(\xi)$ devive from eq.~(48). For the SM one has $\eta_{0,SM}\approx0.6$ (see Table 1). In the case of a deflagration, eqs.~(45,46) have to be used to connect the bubble wall solution to the shock front.

In the last part of this work we have applied our approach to the SM augmented with a $\phi^6$ operator in the Higgs potential. For $\xi\approx1$, i.e. a phase transition strong enough to prevent baryon number washout, we find a wall velocity $v_w=0.3-0.4$, i.e.~subsonic walls, which support the standard picture of electroweak baryogenesis assumed in \cite{Bodeker:2004ws}. As the strength of the dimension-6 operator increases, the phase transition gets stronger and the walls become faster and at some point become supersonic (see Fig. 9). Close to the speed of sound our computation is not reliable, as in the evaluations we linearize in the plasma velocity. For very strong phase transitions, $\xi>3.3$ we observe runaway behaviour of the bubble wall, consistent with \cite{Bodeker:2009qy}. 

It is interesting to note that ref.~\cite{Bodeker:2009qy} provides a direct criterion for runway behaviour. We can use this to fix $\eta_{0,SM}$ independently of \cite{Moore:1995si}. The two determinations of the friction parameter agree well, which is very satisfactory (see Table 2). We take this as additional support for the method presented in this work.

It will be very interesting and fruitful to use this formalism to compute the wall velocities in other promising extensions of the SM, such as the Two-Higgs-doublet model or non-minimal SUSY models to arrive at more reliable predictions for relics of a first-order electroweak phase transition.

\section*{Acknowledgements}
 We like to thank Mark Hindmarsh and Thomas Konstandin for valuable discussions. SH acknowledges support from the Science and Technology Facilities Councel [grant number ST/G000573/1].

\end{document}